# Deeper detection limits in astronomical imaging using self-supervised spatiotemporal denoising

Yuduo Guo[1,3†], Hao Zhang[1,3†], Mingyu Li[2†], Fujiang Yu[2], Yunjing Wu[2,4], Yuhan Hao[1,3], Song Huang[2], Yongming Liang[5,6], Xiaojing Lin[2], Xinyang Li[7], Jiamin Wu[1,3*], Zheng Cai[2,8,9*], Qionghai Dai[1,3*]

[1]Department of Automation, Beijing National Research Center for Information Science and Technology, Tsinghua University, Beijing, China

[2]Department of Astronomy, Tsinghua University, Beijing, China

[3]Institute for Brain and Cognitive Sciences, Tsinghua University, Beijing, China

[4]Kavli Institute for the Physics and Mathematics of the Universe (WPI), The University of Tokyo, Kashiwa, Japan

[5]National Astronomical Observatory of Japan, Tokyo, Japan

[6]Institute for Cosmic Ray Research, The University of Tokyo, Kashiwa, Japan

[7]College of AI, Tsinghua University, Beijing, China

[8]School of Mathematics and Physics, Qinghai University, Xining, China

[9]Tsinghua Wuxi Institute of Applied Technology, Wuxi, China

[†]These authors contributed equally to this work

[*]Corresponding authors: qhdai@tsinghua.edu.cn (Q. D.), zcai@tsinghua.edu.cn (Z. C.), wujiamin@tsinghua.edu.cn (J.W.)

**The detection limit of astronomical imaging observations is limited by several noise sources. Some of that noise is correlated between neighbouring image pixels and exposures, so in principle could be learned and corrected. We present an astronomical self-supervised transformer-based denoising algorithm (ASTERIS), that integrates spatiotemporal information across multiple exposures. Benchmarking on mock data indicates that ASTERIS improves detection limits by 1.0 magnitude at 90% completeness and purity, while preserving the point spread function and photometric accuracy. Observational validation using data from the James Webb Space Telescope (JWST) and Subaru telescope identifies previously undetectable features, including low-surface-brightness galaxy structures and gravitationally-lensed arcs. Applied to deep JWST images, ASTERIS identifies three times more redshift ≳ 9 galaxy candidates, with rest-frame ultraviolet luminosity 1.0 magnitude fainter, than previous methods.**

The imaging detection limit in observational astronomy determines how faint an object can be observed with a given instrument and exposure time (*1–3*). It is limited by noise from the instrument (*4–7*), sky background (*8*), and photon statistics (*9*). Larger-aperture telescopes and advanced instrumentation can collect more photons, reduce instrumental readout noise, and thus improve the signal-to-noise ratio (S/N) of the resulting image (*10–12*). However, other sources of background noise— such as those from zodiacal emission, Galactic foregrounds (*13*), and atmospheric scattering (*8*)—cannot be corrected by those advances in hardware.

Software signal processing algorithms have been used to improve the S/N in astronomical imaging. A widely adopted method involves aligning and co-adding multiple exposures. If there are $M$ exposures, each in the background-dominated regime, this theoretically improves the S/N by a factor of $\sqrt{M}$ (Fig. 1A), assuming pixel-level noise is independent and identically distributed (i.i.d.) (*14–16*). However, this approach provides diminishing returns for deep surveys, because the required exposure time scales prohibitively with increasing depth. Taking the James Webb Space Telescope (JWST) as an example, increasing its 5σ (standard deviation) detection limit from 31 to 32 magnitudes (mag, on the AB system unless otherwise stated; higher numbers are fainter) in the F115W band (this notation indicates a wideband filter centered at 1.15 μm) would require increasing the total exposure time from 4 days to approximately one month (*17*). Convolution with a smoothing kernel can also improve the detection limit (*18*, *19*), by reinforcing local pixel correlations and suppressing high-frequency noise. However, smoothing methods inevitably degrade the spatial resolution.

Deep learning algorithms have been employed in denoising techniques for astronomy (*20–23*). Supervised deep denoising methods utilize machine learning from pairs of noisy and clean images, from single (*24*) or multiple exposures (*25*, *26*). In astronomy, such image pairs are typically generated synthetically (*21*) or through paired long and short exposures (*22*, *23*). The effectiveness of these methods is limited by substantial differences between synthetic and real data, coupled with the complexity and variability of noise in astronomical imaging. Collection of long and short exposure pairs is constrained by the limited observing time available, restricting the availability of reliable training data.

Alternative self-supervised methods require only a single image (*20*, *27–29*) or pairs of noisy images (*21*) for training; several of these methods are based on the Noise2Noise (N2N) algorithmic framework (*30*). These methods typically rely on the overall image S/N as the primary performance metric, which does not necessarily describe the practical improvement for astronomical applications. These methods can introduce artefacts and miss real sources after processing, even in images that appear to be visually clean (*31*).

To mitigate cosmic ray hits, pixel saturation, and spatially varying background emission, astronomical imaging observations typically adopt multiple dithered exposures (*32*). Co-addition (also known as stacking) methods average pixel values across the multiple exposures, while rejecting outlier pixel values (Fig. 1A). However, the assumption of co-addition that noise is pixel-level i.i.d. is generally violated in real data, due to spatial correlations introduced by the telescope's point spread function (PSF), structured background emission, and instrumental effects. Therefore, in practice co-addition methods do improve the S/N, but a deeper detection limit remains theoretically achievable. Algorithms that capture those spatial correlations could further improve the performance after co-addition. For example, N2N-based methods (*30*) employ convolutional neural networks (CNNs) to capture local spatial correlations in a single co-added image, providing improved noise suppression (Fig. 1A).

**Concept, design, and implementation**

We extended the N2N concept of single-image denoising, to multiple exposures of the same target field. We designed a machine learning algorithm, which we name ASTERIS, to exploit the spatiotemporal correlations between multiple exposures, with the goal of producing a higher-S/N image during the co-addition process of a stack of exposures. The algorithm is applied after conventional imaging data reduction but before the identification of sources in the image (Fig. 1A).

ASTERIS employs a tailored spatiotemporal learning strategy, using a dedicated attention mechanism (the detailed neural network architectures are shown in Fig. S1A) to fuse structured voxel (i.e., three-dimensional neighboring pixels sampled across multiple exposures) information in exposures that are astrometrically aligned to the same world coordinate system (WCS).

For self-supervised training, ASTERIS requires a total of 16 exposures of the same field, which are divided into two independent sets: an input set and a target set, each containing eight exposures. Because these exposures are astrometrically aligned, both sets share the same underlying signal expectation, while their noise realizations remain independent (*33*). By minimizing the loss function, the neural network learns to estimate the signal expectation of the input exposures using the target exposures as a clean-reference proxy. The loss function is a composite of two components: (a) the Mean Squared Error (MSE) between the co-addition of the input set and the target set (referred to as average loss), and (b) the Mean Absolute Error (MAE) calculated between individual corresponding exposures from each set (referred to as average loss). We find that this approach reduces complex background noise, preserves faint astronomical signals that would otherwise be buried beneath local noise fluctuations, and reduces false positive detections (*33*). Once trained, the algorithm can perform denoising on any eight input exposures as used in JWST deep imaging surveys (*32*), without requiring additional reference data. We anticipate that this algorithmic framework can also accommodate different numbers of input exposures (as discussed below).

Given the wide dynamic range of most astronomical imaging observations, and because noise primarily affects faint sources, ASTERIS selectively operates on the fainter parts of each image. Specifically, we impose an input flux threshold of 3σ significance above the background noise, chosen for practical reasons (*33*). Pixels below this threshold are retained for denoising by the neural network, while pixels above 3σ are temporarily clipped, median combined, and later directly re-integrated into the final image. This approach preserves the original dynamic range without introducing a discontinuity in flux values (Figs. 1B-D, S2, S3). Imposing this threshold concentrates the neural network's learning capacity on low-S/N parts of the image, thereby optimizing for faint-source detection and characterization.

**Algorithmic performance**

To quantitatively evaluate the performance of ASTERIS, we developed an evaluation pipeline using mock data, an established method in observational astronomy. To generate the mock data, we utilize real images in the F115W filter taken from the JWST gravitational lensing & NIRCam imaging to probe early galaxy formation and sources of reionization (GLIMPSE) program (*34*), comprising 168 dithered exposures at a single pointing. Mock sources are injected into the real data (Table S2), and then we test their recovery using ASTERIS. This approach ensures the independence and variability of background realizations.

The images were resampled to a pixel scale of 0.04″ (arcseconds), then a source-free 128 × 128 pixel region was selected as the background. We generate 2000 independent background realizations by randomly selecting sets of eight exposures from the 168 available images ($\approx 6 \times 10^{15}$ possible combinations, Fig. S4A). This ensures that the set of realizations are statistically independent and free from sampling bias. Astronomical observations typically assess their completeness for point sources, so we inject 25 isolated mock point sources in each background realization, for a total of 50,000 mock sources. The assumed distribution of mock source properties was a cubic power-law from 27.5 to 31.5 mag (Fig. S4B). For the mock testing, we used an ASTERIS model trained on real JWST data composed of program IDs 3293, 1215, 3210, and 1963 (*33*), which converged in approximately 10 epochs (~ 26 hours). Inference on each 8-exposure $1650 \times 1650$ pixel image stack required ~ 18.1 seconds to generate a denoised $1650 \times 1650$ pixel output image, using 4 Graphics Processing Units (GPUs, 40 GB GPU memory each). Detailed computational costs are discussed in the supplement (*33*).

To assess the performance of ASTERIS (trained on JWST data, Table S1), we compare the results with existing methods including co-addition (outlier-rejected averaging), Gaussian smoothing (*18*), Block-Matching and 3D filtering (BM3D) (*35*), Block-Matching and 4D filtering (BM4D) (*36*), Neighbor2Neighbor (*27*), and N2N (*30*) (Fig. S5). We used the version of N2N implemented using the Restormer architecture (*37*), similar to that used in ASTERIS. Sources were identified in the output image produced by each method using the `Source Extractor` software (*38*), in an identical configuration (*33*). We find that ASTERIS consistently yields narrower image histograms (Fig. 2B-D, which compares ASTERIS to co-addition and N2N), indicating lower noise standard deviation, and more true positive sources (as shown in Fig. 2D, the value $n$ in the bottom-right corner of each panel denotes the number of identified true positives). We calculate the source-dependent photometric S/N using circular apertures of 0.14″ radius, which enclose approximately 80% of the PSF flux in these JWST data (Fig. S4C). Using this metric, the resulting $5\sigma$ sensitivity provided by N2N is an improvement over co-addition, and ASTERIS is an improvement over N2N (Fig. 2E).

The photometric S/N, which quantifies visual performance, does not fully capture the practical improvements in astronomical applications. We performed an analysis of source completeness (the fraction of sources correctly detected) and purity (the fraction of detections that are true positive) (*33*), which are commonly employed to quantify detection performance in astronomical imaging. At the 90% completeness level, N2N improves the detection limit by 0.1 mag compared to co-addition, while ASTERIS improves it by about 1.0 mag (Fig. 2F). At 90% purity, ASTERIS improves by more than 1.5 mag compared to co-addition, and 1.0 mag compared to N2N, indicating a concurrent reduction in false positives (Fig. 2G). When evaluated using the F-score metric (equation S7) (*33*), which combines completeness and purity, ASTERIS improves the limiting depth (at F-score = 0.9) by about 1.7 mag compared to co-addition (Fig. 2H) and 1.4 mag compared to N2N (Fig. S5K). We find similar improvements under less controlled conditions, with a larger field and bright sources masked out (Fig. S6), which more closely resemble real observations. We also performed quantitative analysis of the performance variations resulting from different combinations of loss functions during training (Fig. S7) and different flux thresholds for sigma-clipping methods used in ASTERIS (Fig. S8).

To investigate the effect of ASTERIS on spatial resolution and photometric measurements, we evaluated the PSF fidelity and photometric accuracy (*33*). We derived the PSF profiles output by each method using the radially averaged flux around the extracted PSF centers (Fig. 2I). A two-sample Kolmogorov–Smirnov test (*39*) comparing the PSF profiles from ASTERIS and co-addition yielded a $p$-value of 0.9, indicating no statistically significant difference between them.

In contrast, the same test for N2N yielded a statistically significant $p < 0.05$, which we interpret as indicating degraded resolution in the N2N output. We performed additional tests of photometric accuracy (*33*), finding no evidence of systematic biases introduced by ASTERIS, and slightly improved photometric precision for faint sources, compared to co-addition (Fig. 2J). We also verified that the ASTERIS output is robust to the permutation order of the input exposures (Figs. S9-S10).

**Observational validation and generalization**

Detecting faint sources or low-surface-brightness features (*40*) is challenging in astronomical imaging. We evaluate the performance of ASTERIS for this task, using the same trained model as in Fig. 2 (*33*). We use space-based JWST data and ground-based data from the Subaru telescope to assess the generalization capability. Note that all the data used for testing is excluded from the training dataset (Table S1).

We again select NIRCam data from the JWST GLIMPSE program (*34*): eight exposures (independent from the training data) for co-addition benchmarking and ASTERIS denoising, and a deep co-addition of 168 exposures serving as ground truth. For this test, ASTERIS utilized not only information derived from the eight exposures, but also the statistical priors previously learned from the training datasets (Table S1), which used the same instrument. Filters involved in this test are F115W (Fig. 3A-F, J-L), F090W (Fig. 3G-I), and F444W (Fig. 3M-O).

For faint sources, especially those in a crowded field near bright sources (*41*), the presence of spatially complex background noise compromises both source detection and accurate flux estimation (Fig. 3A, S11A). We analyze a representative crowded field containing numerous faint sources that were undetected in a standard co-addition of the 8-exposure NIRCam stack, but were detected in the 168-exposure co-addition of the same field. We identify 168 sources in the 168-exposure image using `Source Extractor` (Figs. 3C, S11C), of which 93 were recovered by ASTERIS from the 8-exposure input (Fig. S11B). We verified that the cross-matched sources are true positives and preserve photometric accuracy (Fig. 3B-F). In contrast, the conventional 8-exposure co-addition recovers only 50 of these sources (Fig. S11A) under identical `Source Extractor` settings. When applied directly to the full 168 exposures, ASTERIS leads to the detection of 229 sources (Fig. 3D, S11D), 35% more than were identified using standard methods. This demonstrates that ASTERIS can identify additional sources in an existing dataset, such as the imaging data from program ID 4111 (Fig. S12) (*42*) — without the need for additional exposure time or hardware upgrades.

To investigate low-surface-brightness features, we consider the extended stellar disks and diffuse outer arms of spiral galaxies (*43*, *44*), which are often undetectable in observations (Fig. 3G to I, S13B). We apply ASTERIS to F090W data from the JWST GLIMPSE program and calculate the structural similarity (SSIM) metric (*45*), which quantifies the suitability of an image for morphological characterization of galaxies. For an 8-exposure co-addition, we find that ASTERIS improves the SSIM value from 0.37 (Fig. 3G) to 0.67 (Fig. 3H), compared to standard methods. In the same image, we find that ASTERIS also improved the reconstruction of a low-surface-brightness gravitationally lensed arc (Fig. 3J-L, S13D), the distorted image of a distant galaxy (*46*). For this lensed arc, the SSIM increases from 0.59 (Fig. 3J) using standard co-addition to 0.81 (Fig. 3K) using ASTERIS. We find similar improvements in the identification of groups of faint and diffuse galaxies (Fig. 3M-O, S13F).

To investigate whether ASTERIS can generalize to data from different instruments, we utilize observations from the ground-based Subaru telescope (8.2-meter aperture) using the Multi-Object Infrared Camera and Spectrograph (MOIRCS) (*47*, *48*). Compared to the space-based JWST data, these observations have lower spatial resolution resulting from the atmospheric turbulence (seeing), and subject to higher sky background noise (Fig. 3P). Directly applying the JWST-trained model leads to some false positive sources (Fig. S14B). Consequently, we trained a dedicated ASTERIS model using Subaru datasets from programs S17A-198S (*49*), S24A-091, and S24B-135 (*33*). Then, we performed a similar test to the JWST data, applying ASTERIS to a stack of eight 3-second Subaru exposures, and comparing the results to a deeper (2088 seconds) co-added exposure of the same field (Fig. 3R) using the same instrument in the K-short (Ks) band (a near-infrared filter centered at 2.15 μm). We find that ASTERIS recovers faint sources in this ground-based data (Fig. 3Q, S13H).

**Application to faint high-redshift galaxies**

As an example application of ASTERIS, we search for high-redshift galaxies, the most distant and earliest known galaxies with extremely faint apparent magnitudes. We employ ASTERIS to process one of the deepest JWST imaging datasets, the JWST Advanced Deep Survey (JADES) Origins Field (JOF) (*50*, *51*). This program comprises 14 filters from F090W to F444W; a detailed introduction of the data is presented in Table S1 (*33*). Previous works have identified a population of high-redshift galaxy candidates in the JOF, with redshifts $z \sim 9$ to 15 (including both photometric redshift $z_{\mathrm{phot}}$ and spectroscopically confirmed redshift), with the faintest having rest-frame absolute ultraviolet (UV) magnitudes $M_{\mathrm{UV}} = -17$ (*52*, *53*). We run ASTERIS on the JOF data, identify sources using `Source Extractor`, and select high-redshift galaxy candidates using consistent criteria with the previous work (*33*, *53*).

We find the 5σ depth of the JOF across all bands improves from 30.3 to 30.8 mag in previous work (*53*), to 30.9 to 31.6 mag using ASTERIS (*33*) (Fig. S15). We identify 162 high-redshift galaxy candidates at $z_{\mathrm{phot}} \gtrsim 9$, approximately three times the number previously identified (Table S3, Figs. 4, S16). The faintest candidates we identify have $M_{\mathrm{UV}} \approx -16$ mag, one magnitude fainter than in previous work. All the high-redshift galaxies in JOF that have been spectroscopically confirmed by previous work (*1*), or identified as candidates by multiple previous works (*52*, *53*), were recovered using ASTERIS. 75% of the high-redshift galaxy candidates identified using ASTERIS were not reported by previous works (*52*, *53*). Table S3 shows the measured $M_{\mathrm{UV}}$ and photometric redshifts for all the candidate sources. We assessed the detection and selection completeness using source-injection simulations (Fig. S17) (*33*).

We ascribe this threefold expansion in the number of candidates identified to two primary effects of ASTERIS. Firstly, ASTERIS enables the detection of fainter sources, with apparent magnitude of 31 to 32 (Fig. 5A-B). By suppressing the noise, ASTERIS improves the detection limit allowing previously marginal signals to emerge as statistically significant detections. Secondly, the improved S/N for faint sources helps to resolve photometric redshift ambiguities (Fig. 5C-D). High-redshift galaxy candidates are selected using the Lyman-break technique (*54*), in which intervening neutral hydrogen in the intergalactic medium absorbs UV light blueward of the Lyman-α line (1215.6 Å). The wavelength of this break is redshifted into the near-infrared for $z \gtrsim 9$ objects, causing them to appear bright in longer-wavelength filters but invisible (a "dropout") in shorter-wavelength filters. Therefore, a strong break or non-detection ($S/N < 2$) of galaxies at wavelengths shorter than the rest-frame 1215.6 Å serves as the primary indicator of a high-redshift source. However, some low-redshift galaxies have a similar break at the shorter rest-frame wavelength 0.4 μm due to the Balmer break (*55*, *56*). If the brightness of those sources is similar to the noise level, a low-redshift Balmer break can be misidentified as a high-redshift Lyman break, or the data are unable to distinguish between the two possibilities. ASTERIS reduces this effect by providing deeper detection limits in the dropout bands, thereby better constraining the source photometric redshift, particularly for sources that were marginally significant (S/N ~ 2) using standard co-addition methods (Fig. 5D).

The JOF field contains some of the highest-redshift galaxy candidates ($z > 16$) haven't been identified by previous works (*52*, *53*). ASTERIS identified four of these candidates at $z \sim 16$ to 22.5 in total (Table S3). An example is shown in Fig. 5E, where the source is a dropout in the F200W band. The redshift probability distribution derived using ASTERIS (Fig. S18A) shows the high-redshift solution ($z_{\text{phot}} = 17.08$) is far more probable than the alternative low-redshift solution (Fig. S18B), resulting in a more secure identification.

We use the identified galaxy candidates to quantify the rest-frame UV luminosity functions, which describe the number density of galaxies with different luminosities in selected redshift bins. The ASTERIS analysis identified 125 galaxy candidates at $z \sim 9$ to 12 and 33 at $z \sim 12$ to 16, within an observed volume that is substantially smaller than previous studies (*52*, *53*). The resulting UV luminosity functions (Fig. 6), fitted with models, have steep faint-end slopes of –2.45 ± 0.03 for $z \sim 9$ to 12 and –2.28 ± 0.02 for $z \sim 12$ to 16 (*33*). These values indicate higher number densities of faint galaxies down to $M_{\text{UV}} \sim -16$ than predicted by theoretical models (*57*, *58*).

**Extension to different numbers of exposures**

Due to the neural network architecture (Fig. S1), ASTERIS only supports an even number of input exposures. Therefore, the 8-exposure input strategy employed in ASTERIS was chosen to align with the recommended observing strategy for deep JWST surveys (typically 9-point dithers) (*32*) and to balance the neural network's memory requirements and computational efficiency. To test its applicability to other survey strategies, we also developed a 4-exposure variant of ASTERIS (*33*). Compared to the 8-exposure version, the 4-exposure neural network provides a smaller improvement in detection limit—from 1.0 mag to 0.7 mag at 90% completeness—with a corresponding decrease in the identification of faint sources (Fig. S19). We attribute this degradation to the reduced availability of spatiotemporal information in the smaller number of exposures.

We strongly discourage artificially duplicating exposures to increase the number of input exposures, because this would violate the assumptions required for the spatiotemporal learning strategy. Tests of this approach show it led to an increase in the false positive rate (Fig. S20). In the opposite situation, where more exposures are available than accepted by ASTERIS, they can be grouped into subsets and each subset combined via co-addition to produce the required number of exposures. For example, a set of 24 exposures can be co-added in eight subsets of three exposures each, then the eight combined exposures are used as input for ASTERIS (Fig. S21A). We also tested situations where the number of available exposures is not an even multiple of those used by ASTERIS, such as 21 exposures used by the 8-exposure version of ASTERIS (Fig. S21B), or seven exposures used by the 4-exposure version of ASTERIS (Fig. S21C).

## Methods summary

We used JWST NIRCam imaging data under program IDs 3293 (*34*), 1210, 3215 (*50*, *51*), 1963 (*59*), and 4111 (*42*), spanning filters from F070W to F480M (this denotation indicates a medium band filter centered at 4.80 μm), for the training and testing of ASTERIS. We conducted data reduction using the JWST Science Calibration Pipeline (`jwst` v1.15.1 (*60*)) and additional procedures to mitigate the impact of various artifacts (*61, 62*). We utilized public catalogs (*50*, *63*) for astrometric correction of all individual exposures. We combined individual exposures (with the same astrometric pointing) along the temporal domain to form a 3D data cube, and then applied a 3σ clipping to the data cube to separate it into a faint part and a bright part; only the latter was retained for ASTERIS training. Then, the faint part was Z-score normalized (equation S4) prior to being input into the neural network.

The neural network architecture of ASTERIS utilizes a spatiotemporal 3D U-Net framework enhanced by multi-deconvolved-head-transposed-attention to overcome the receptive field limitations of traditional CNNs. We trained ASTERIS separately for the long-wavelength (filter central wavelength ≥ 2.5 μm) and short-wavelength (< 2.5 μm) NIRCam imaging data. The loss function comprised a weighted combination of MSE and MAE losses (equations S1-S3). The training dataset was segmented into 120,000 patches (60,000 pairs) with each patch formatted as an 8 × 128 × 128 pixels cube. The training reached convergence in approximately 10 iterations (~26 hours). For denoising, ASTERIS partitions input exposures into patches according to the specified patch size and the number of available GPUs. We benchmarked ASTERIS against previous denoising methods using mock tests on the JWST NIRCam F115W data, where the mock sources were generated using the model PSF generated from the `STPSF` software (*64*). These mock sources, ranging from 27.5 to 31.5 mag, followed a third-order power-law distribution to match the source counts in the real observations. We quantified the performance using metrics including completeness, purity, and the F-score (equation S7, (65, *66*)). Then, we extracted the corresponding PSF from different methods using the `PSFEx` software (*67*) and analyzed their profiles by radial averaging. Photometric accuracy was further validated through forced-aperture photometry at known mock-source locations.

The trained ASTERIS was then applied to identify faint high-redshift galaxy candidates within the JOF (*50*, *51*). Source detection was performed using `Source Extractor` (*38*) on a composite detection image generated by co-adding the ASTERIS-denoised images in F277W, F356W, and F444W. We conducted forced-aperture photometry with a 0.1″ radius circular aperture across all filters, applying aperture corrections based on PSF profiles and the flux ratio between circular and Kron photometry (*68*). The photometric uncertainties were estimated with random aperture sampling (same radius as used in photometry) on source-free regions. Consequently, we adapted the criteria from previous work (*53*) for high-redshift galaxy candidate selection, followed by a preliminary spectral energy distribution (SED) fitting using the `EAZY` software (*69*), with a template suite presented in (*70*). We then inferred the physical properties of the candidates with the Bayesian Analysis of Galaxy SEDs (`BEAGLE`) software (*71*) with models detailed in previous work (*53*). Finally, we evaluated luminosity functions (LFs) by assessing completeness through mock source injection and recovery tests. Binned LFs were calculated using the effective volume method (*72*) (equations S8-S9) and characterized by fitting a Schechter function (*73*).

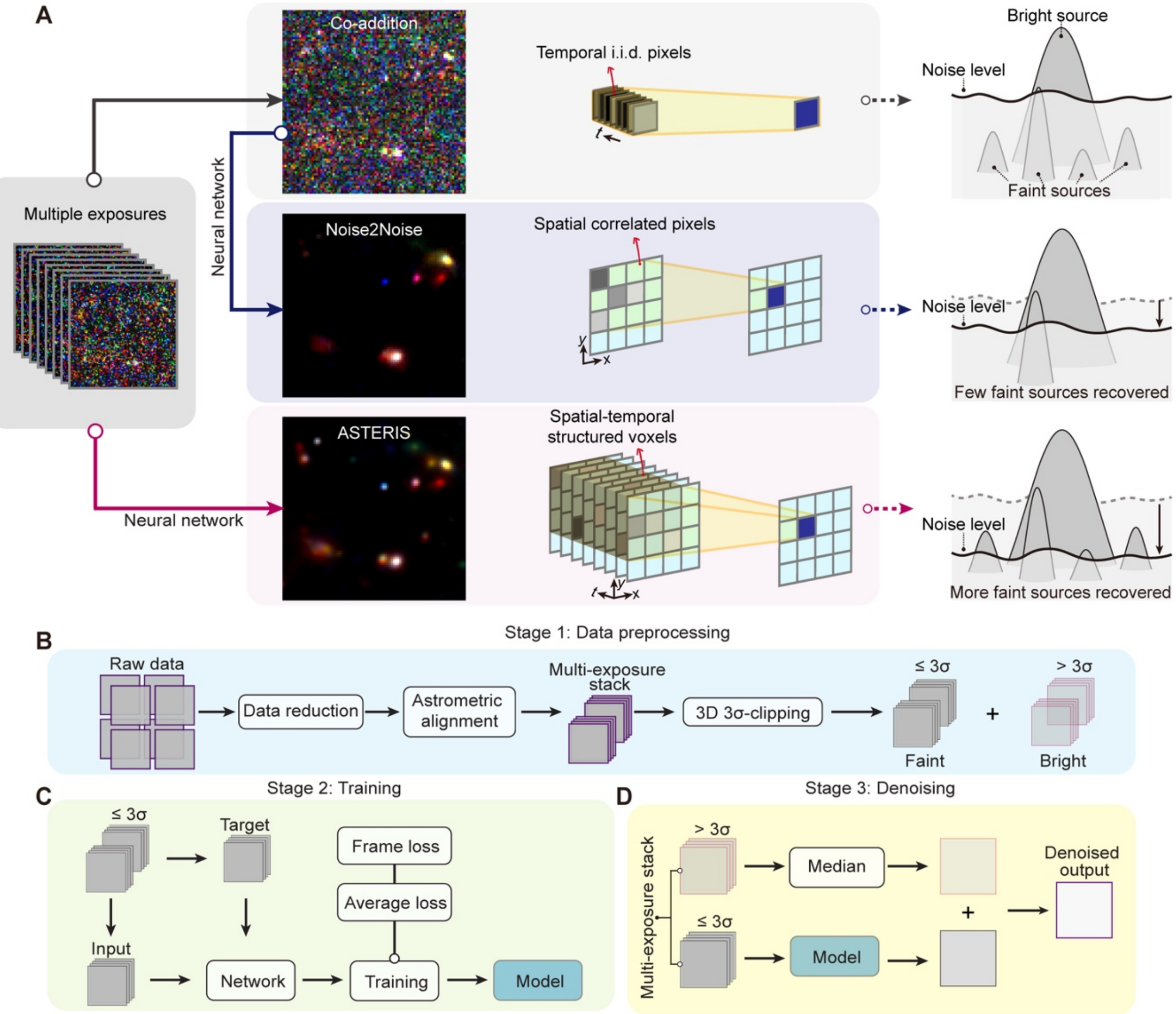

**Fig. 1. Schematic overview of the ASTERIS algorithm. (A)** Comparison between standard co-addition (outlier-rejected averaging), N2N denoising, and ASTERIS denoising. The workflow is organized into four columns from left to right: input data, denoising methods, underlying principles, and conceptual denoising effect. Solid and dashed lines denote data flow and corresponding effect (gray for co-addition, blue for N2N, and red for ASTERIS), respectively. In the third column, multi-colored grids illustrate the input noisy pixels across spatial ($x$, $y$) and temporal ($t$) dimensions, while the blue grid represents the resulting denoised pixel. Co-addition method (top row within gray shadings) increases the S/N by averaging multiple exposures under the assumption of independent pixel-level noise; however, faint signals remain below residual noise fluctuations (standard deviation, std). The N2N deep-learning-based single-frame denoising method (middle row within blue shadings) reduces the noise std (standard deviation) by modelling local spatial correlations, improving the S/N of bright sources (compared to co-addition) but recovers few additional faint sources. ASTERIS (bottom row within red shadings) jointly processes multiple exposures, producing a single denoised frame by learning spatiotemporal correlations. This both reduces the noise std and recovers a larger number of previously undetectable faint sources. **(B-D)** Three stages of the ASTERIS pipeline. In the first stage, pre-processing (B), multiple exposures of the same pointing are combined after data reduction and astrometric alignment, then separated into a bright part and a faint part, using a 3σ clipping threshold. In the second stage, training (C),

the faint part ($\leq 3\sigma$) is randomly sampled into input and target sets for self-supervised learning by a neural network. The model is trained to minimize both an average loss (MSE) and a frame loss (MAE) between the neural network output and the target exposures. In the third and final stage, denoising (D), the faint part ($\leq 3\sigma$) of the multi-exposure image stack is denoised by ASTERIS, then recombined with the co-added median of the bright part ($> 3\sigma$), to produce the final denoised image with full dynamic range.

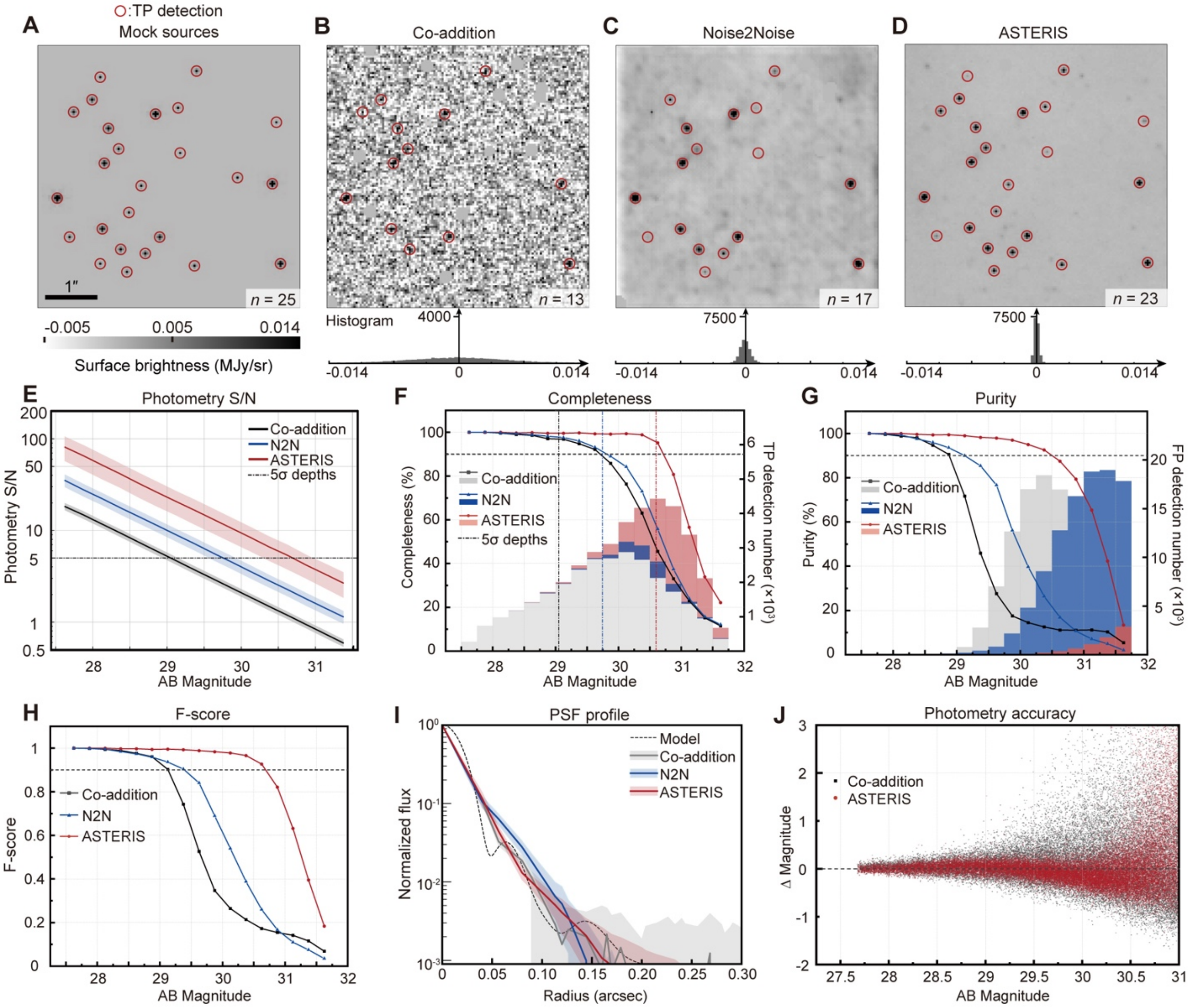

**Fig. 2. Characterization of ASTERIS using mock sources. (A)** Example isolated ground-truth sources (within the red circles, TP means true positive) with a clean background. The black scale bar is 1.0 arcseconds. The resulting images produced by (**B**) co-addition by averaging 8 exposures with mock sources in panel A injection into JWST NIRCam F115W images (grayscale, see color bar in megajansky per steradian (MJy/sr)), (**C**) N2N applied after the same 8-exposure co-addition, and (**D**) ASTERIS run on the 8 exposures. The labelled value *n* in the lower-right corner of each panel indicates the number of sources detected using `Source Extractor` with identical parameters. Histograms below panels B-D show the corresponding pixel values, as a measure of the background noise variance. **(E)** Colored lines show the derived source-dependent photometric S/N after denoising by each method (see legend). Solid lines show the mean S/N, binned by 0.25 mag, from 50,000 injected sources; shaded regions indicate the ±1σ scatter. The dot-dash black line in panels E to F indicates the 5σ detection threshold. **(F)** Detection completeness (lines with data points) and number of true positives (histograms) for each method. The 90% completeness level is plotted by a black dashed line. **(G)** Same as panel F but for detection purity (lines) and number of false positives (histograms). The black dashed line shows the 90% purity level. **(H)** The F-score metric for each method. The black dashed line indicates an F-score of 0.9. **(I)** The PSF profiles from each method, using 0.04″ pixel sampling. The model PSF is generated from the STPSF Python Package (*64*), sampled at 0.01″. **(J)** Comparison of photometric accuracy for the 50,000 mock sources, as measured using co-addition (black points) and ASTERIS (red points). The dashed black line indicates no difference between the mock injection and the derived measurement. The data in panels E, F, and J are all measured using 0.14″ radius circular apertures (*33*). Fig. S6 shows another example in a more crowded and larger field.

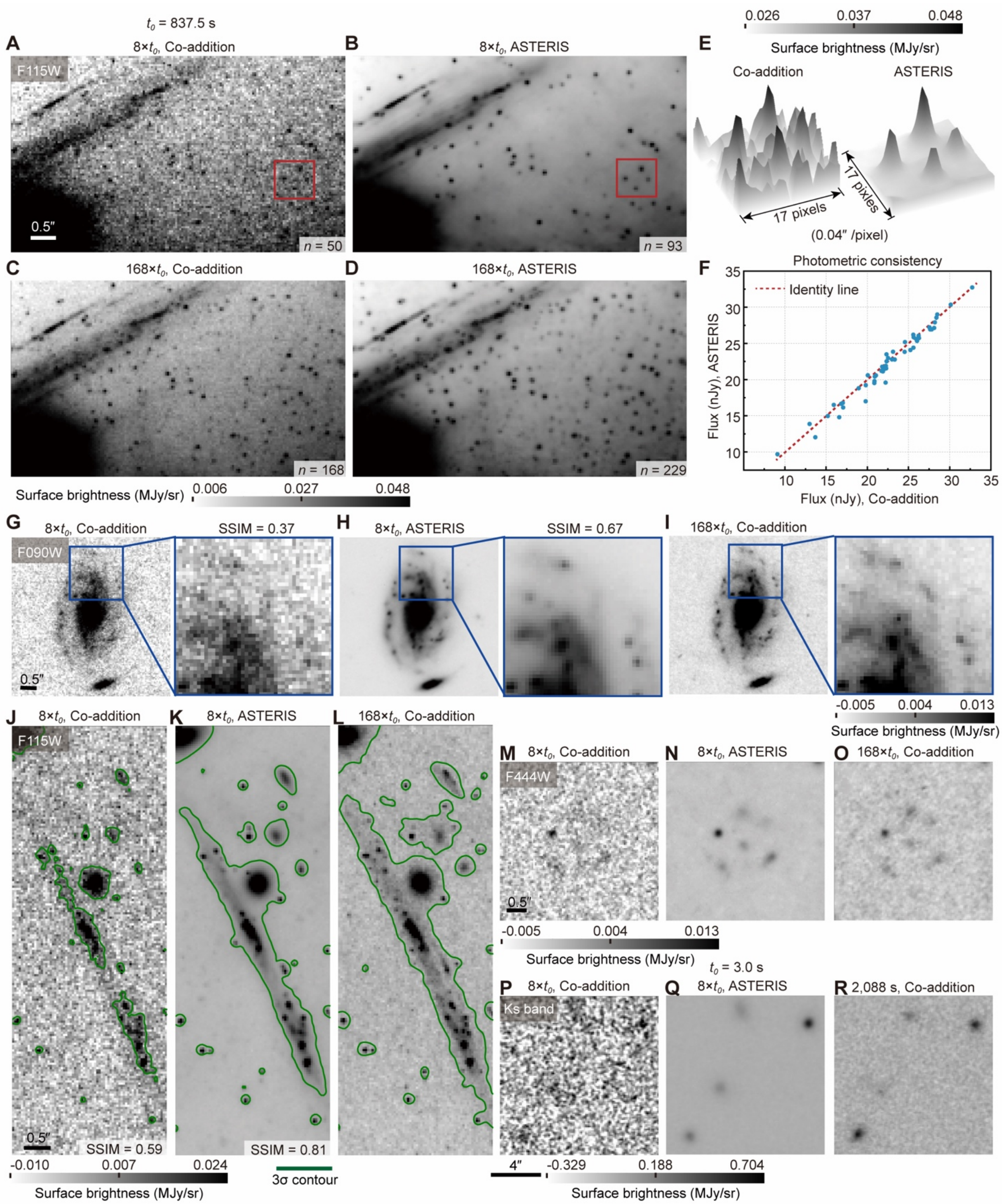

**Fig. 3. Observational validation of ASTERIS on real data.** $t_0$ is the individual exposure time and $M\times$ indicates that $M$ exposures were used. **(A** to **D)** A crowded field observation from JWST NIRCam in F115W, with $t_0 = 837.5$ s. $n$ is the number of sources identified by `Source Extractor` with fixed parameters. The images resulting from (**A**) co-addition of $8 \times t_0$ exposures, (**B**) ASTERIS applied to $8 \times t_0$ exposures, and (**C**) co-addition of $168 \times t_0$ exposures, serving as the ground truth for panels A and B. (**D**) ASTERIS applied to $168 \times t_0$ exposures, which shows

fainter sources than the ground truth image in panel C. **(E)** 2D surface profiles of the region within the red-boxes in panels A and B. **(F)** Comparison of the measured fluxes for true-positive sources in panels A and B. The dotted red line indicates identical values in the two methods. **(G** to **I)** Similar to panels A-C, but for an example spiral galaxy observed using JWST NIRCam in F090W, with a zoom-in view of the spiral arms. The labeled SSIM values of the zoomed-in regions in panels G and H were computed relative to the image in panel I. **(J** to **L)** Similar to panels G-I, but for an example gravitational lensing arc, observed using JWST NIRCam in F115W. Green contours are at 3σ significance. The labelled SSIM values in panels J and K were computed inside the contours, relative to the image in panel L. **(M** to **O)** Similar to panels J-L, but for an example group of faint and diffuse galaxies observed using JWST NIRCam in F444W. Scale bars in panels A, G, J, and M are 0.5 arcseconds. **(P** to **R)** Similar to panels M-O, but for another group of faint sources observed using Subaru MOIRCS in Ks-band. For these data, $t_0 = 3$ s. (**R**) Co-addition of 2,088 s serves as the ground truth for panels P and Q. Scale bars in panel P are 4 arcseconds. Fig. S12 shows a comparison of spatial profiles taken from the same three example regions.

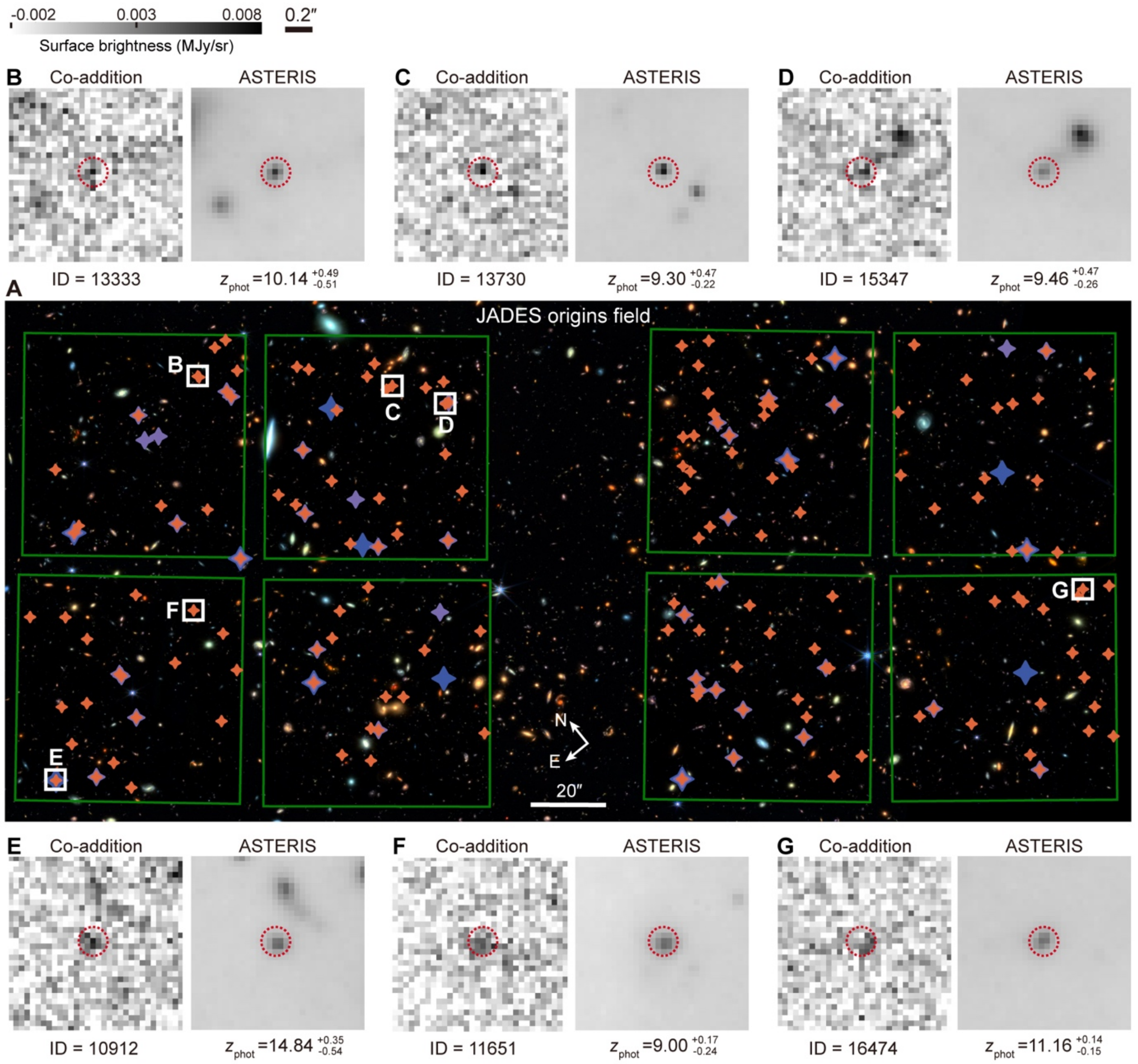

**Fig. 4. Application of ASTERIS to faint high-redshift galaxy candidates in the JADES Origins Field (JOF). (A)** False-color RGB composite image of the JOF NIRCam imaging processed using ASTERIS. Blue is F115W + F150W; green is F200W + F277W; red is F356W + F444W. Green boxes outline the boundaries of the F200W image footprint. The scale bar is 20 arcseconds. Diamond symbols indicate high-redshift galaxy candidates from previous studies [blue (*52*) and purple (*53*)] and using ASTERIS (orange) within the F200W footprint. North and east are denoted. (**B** to **F**) White boxes highlight selected candidates, with zoomed-in thumbnails from the F200W images produced by standard co-addition and ASTERIS. Red dashed circle denotes the candidate sources. $z_{\mathrm{phot}}$ is the best-fitting photometric redshift and ID numbers correspond to entries in Table S3. The scale bar and the diameter of red dashed circles are each 0.2 arcseconds.

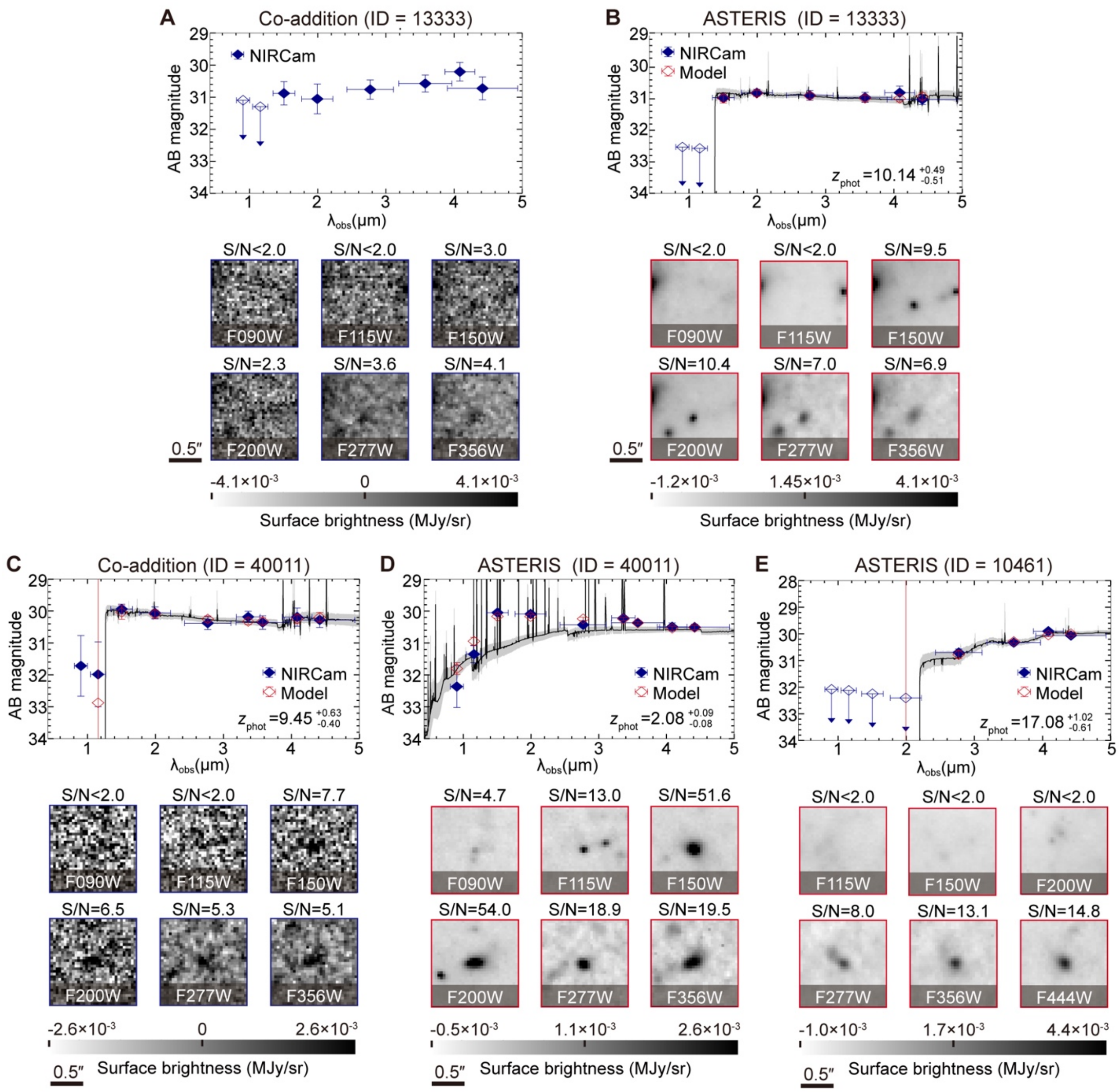

**Fig. 5. Primary effects of ASTERIS in revealing the high-redshift galaxy candidates. (A** to **E).** Each panel plots the spectral energy distribution (SED) (*33*) and thumbnail images in each filter for some example sources. ID number gives the target designation in Table S3. The horizontal axis denotes $\lambda_{obs}$, the observed-frame wavelength expressed in μm. Blue solid diamonds with 1σ error bars are the observed photometry, black lines are model spectra fitted to the observations, gray shading is the 1σ model uncertainty, and red open diamonds are synthetic photometry computed from the model spectra. For filters with S/N < 1.0, 2σ upper limits are plotted as open blue diamonds with downward arrows. **(A** to **B)** Comparison of results from standard co-addition, (**A**) and ASTERIS (**B**) for a faint F115W dropout candidate around the detection limit in the co-addition image. The depth of the F090W and F115W photometry changes by > 1.0 mag, constraining the redshift. The scale bar in the thumbnail images is 0.5 arcseconds. **(C** to **D)** Same as panels A and B, but for a source with an ambiguous redshift. The F090W photometry had S/N < 2.0 in the co-addition and S/N = 4.7 using ASTERIS, changing the best-fitting redshift from $z_{phot}$ = 9.45 to $z_{phot}$ = 2.08. **(E)** Same as panel D, but for an F200W-dropout high-redshift candidate which was identified in the ASTERIS image but not revealed in the co-addition image. Fig. S16 shows five additional examples.

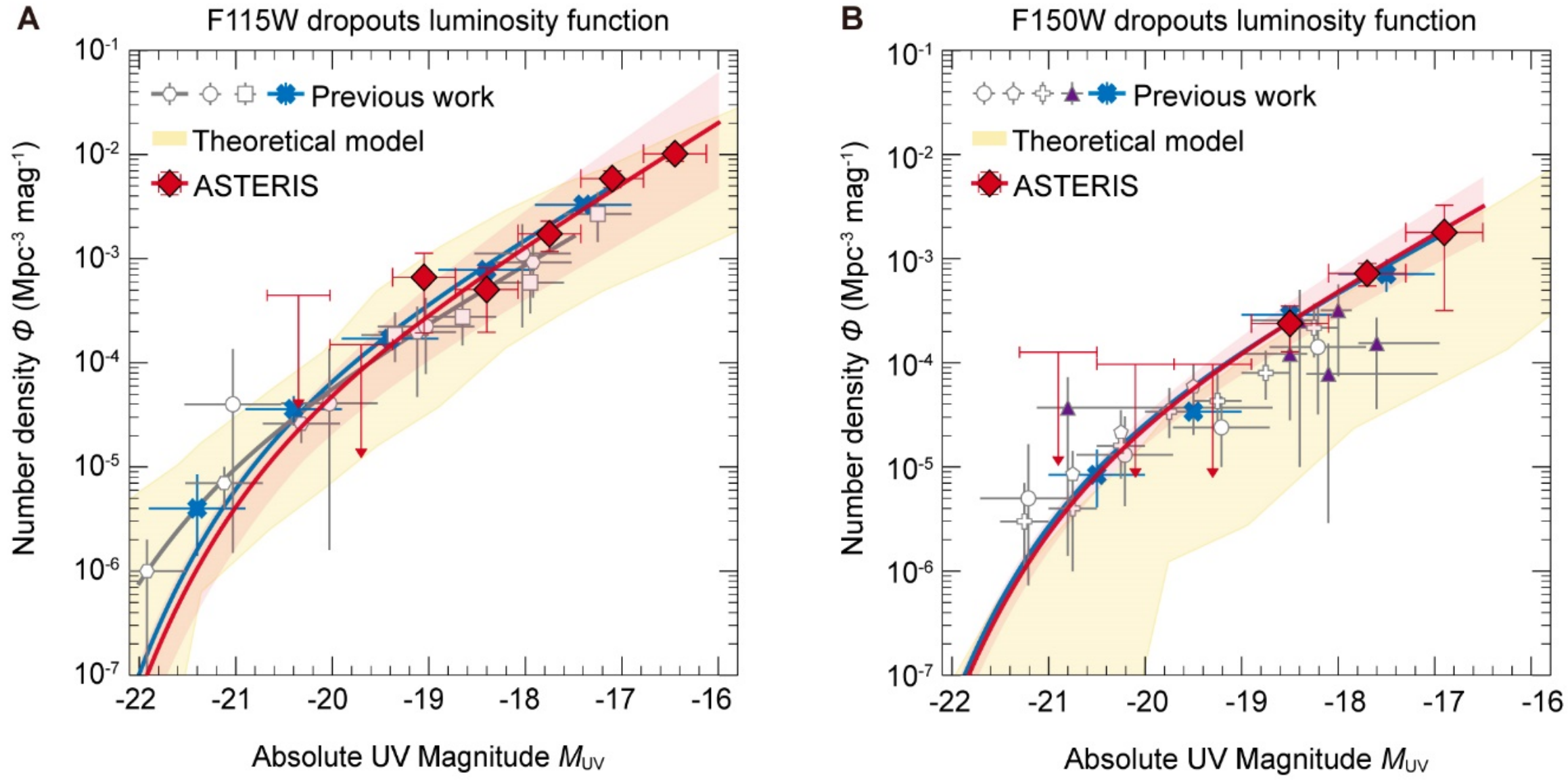

**Fig. 6. Rest-frame UV luminosity functions derived from high-redshift galaxy candidates in JOF. (A** to **B)** The rest-frame UV luminosity functions constrained by the galaxy number density as a function of a given absolute UV magnitude $M_{UV}$. Red diamonds are the binned number densities from the ASTERIS photometry. Red curves are models fitted to the data (*33*), with pink shading indicating their 1σ uncertainties. Blue symbols show equivalent results from a previous analysis covering the JOF using standard co-addition (*53*). Purple triangle symbols present results from another previous JOF study only for F150W dropouts (*52*). Other gray symbols show results from previous studies of other fields [hexagons (*74*), circles (*75*), squares (*76*), pentagons (*77*), and crosses (*78*)]. Yellow shaded regions are theoretical predictions for F115W dropouts (*58*, *79*, *80*) and F150W dropouts (*57*, *79*, *81*). The galaxy number density $\Phi$ has a unit of $Mpc^{-3}$ $mag^{-1}$, where 1 Mpc ≈ 3.09 × $10^{22}$ m. **(A)** F115W dropouts, corresponding to $z \sim 9$ to 12. The ASTERIS image contains 125 galaxy candidates in this range, with a median redshift of 9.51. **(B)** F150W dropouts, corresponding to $z \sim 12$ to 16. The ASTERIS image contains 33 galaxy candidates in this range, with a median redshift of 12.97.

**Acknowledgments:** We thank Chen Zhu, Peixin Weng and Zuyi Chen for helpful discussions. We acknowledge Rhythm Shimakawa as the principal investigator of the Subaru program S17A-198S, and Ichi Tanaka for assistance with the Subaru MOIRCS data processing. We gratefully acknowledge the NASA/ESA/CSA James Webb Space Telescope mission and its public-access data policy, and particularly thank the investigators of programs 1210, 1963, 3215, 3293, and 4111, whose public data were instrumental in the development and fine-tuning of ASTERIS.

This work is based in part on observations made with the NASA/ESA/CSA James Webb Space Telescope. The JWST data were obtained from the Mikulski Archive for Space Telescopes (MAST) at the Space Telescope Science Institute, which is operated by the Association of Universities for Research in Astronomy, Inc., under NASA contract NAS 5-03127 for JWST.

This research is based in part on data collected at the Subaru Telescope and obtained from the SMOKA, which are operated by the National Astronomical Observatory of Japan. We are honored and grateful for the opportunity of observing the Universe from Mauna Kea, which has the cultural, historical, and natural importance in Hawaii.

**Funding:**

Q.D. was funded by National Natural Science Foundation of China, grant number 62088102. Z.C. was funded by National Key R&D Program of China, grant number 2023YFA1605600, National Natural Science Foundation of China, grant number 12525303, and Tsinghua University Initiative Scientific Research Program. J.W. was funded by National Natural Science Foundation of China, grant number 62222508 and 62525506. Y.G. was funded by National Natural Science Foundation of China, grant number 62505156, and China Postdoctoral Science Foundation 2024M761675 and 2025T180226. H.Z. was funded by National Natural Science Foundation of China, grant number 62505157. Y.G., J.W., and Q.D. were funded by Beijing Key Laboratory of Cognitive Intelligence. Z.C. and J.W. were funded by New Cornerstone Science Foundation through the XPLORER PRIZE.

**Author contributions:**

Conceptualization: Y.G., J.W., Z.C.

Methodology: Y.G., H.Z., Y.H., S.H.

Investigation: Y.G., Z.C., H.Z., M.L., Y.W.

JWST data preparation: Z.C., F.Y., M.L., Y.W., H.Z.

Subaru data preparation: M.L., H.Z., Y.L., Z.C.

Luminosity function: M.L., Z.C., Y.G., H.Z., Y.W., F.Y., X.Lin.

Visualization: Y.G., H.Z., M.L.

Funding acquisition: Q.D., Z.C., J.W., Y.G., H.Z.

Project administration: Y.G., Z.C., J.W.

Supervision: Q.D., Z.C., J.W.

Writing – original draft: Y.G., M.L., H.Z.

Writing – review & editing: Y.G., H.Z., M.L., J.W., Z.C., Q.D., S.H., Y.L., X.Lin., X.Li.

**Competing interests:** The authors declare no competing interests.

**Data and materials availability:** The source code for our implementation of ASTERIS in Python is available at https://github.com/freemercury/ASTERIS_THU.git and the version used for this study is archived at Zenodo (*82*) . The demonstration data we used are also archived at Zenodo (*83*). The JWST data are available from MAST https://mast.stsci.edu/portal/Mashup/Clients/Mast/Portal.html under program IDs 1210, 1963, 3215, 3293, and 4111, or can be accessed via DOI (*84*). The Subaru data are archived at Zenodo (*85*). The pre-trained models for JWST NIRCam and Subaru MOIRCS are archived at Zenodo (*86*). No physical materials were generated in this work.

**Supplementary Materials**

Materials and Methods

Figs. S1 to S21

Tables S1 to S3

References *(87–109)*

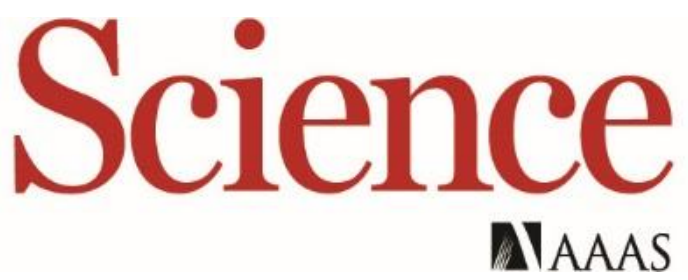

# Supplementary Materials for

## Deeper detection limits in astronomical imaging using self-supervised spatiotemporal denoising

Yuduo Guo *et al.*

Corresponding authors: Qionghai Dai, qhdai@tsinghua.edu.cn;
Zheng Cai, zcai@tsinghua.edu.cn; Jiamin Wu, wujiamin@tsinghua.edu.cn

**The PDF file includes:**

Materials and Methods
Figs. S1 to S21
Tables S1 to S3
References

**Other Supplementary Material for this manuscript includes the following:**

Data S1 (.csv)

# Materials and Methods

## Imaging datasets

### *JWST NIRCam imaging data*

We used near-infrared JWST imaging data under program IDs 3293 (GLIMPSE (*34*)), 1210 and 3215 (JOF (*50*, *51*)), and 1963 (Ultra Deep Field Medium Band Survey (*59*)). Additionally, we used program ID 4111 (Medium bands, Mega Science (*41*)) in ASTERIS testing. Table S1 summarizes the datasets.

The GLIMPSE survey consists of ultra-deep NIRCam imaging of a single field of a massive galaxy cluster, using seven broad-band filters at wavelengths ≥ 0.9 μm, and two medium-band filters, F410M and F480M. Among the filters, F090W, F115W and F444W include 168 independent exposures. This dataset was used for benchmarking of the various denoising algorithms, because individual exposures or subsets can be utilized for network training and interference, while the deep co-added long-exposure images served as ground truth for validation.

The JOF is a deeper single-pointing survey, covering approximately 9 arcmin$^2$ area with 14 filters (7 broad-band and 7 medium-band, see Table S1 for a comprehensive filter list). The number of independent exposures varies from 12 (in F200W and F356W) to 60 (in F182M and F250M). We used the JOF data for an example application of ASTERIS, searching for candidate high-redshift galaxies.

The Ultra Deep Field Medium Band Survey provides deep medium-band imaging with effective exposure times ranging from approximately 1.1 to 7.7 hours in different filters. For the training of ASTERIS, we specifically used the F182M, F210M and F480M filters, each consisting of 24 independent exposures.

The Medium bands, Mega Science survey utilizes two broad-band and 11 medium-band filters. Specifically, the F090W filter comprises 16 independent exposures, while all other bands consist of eight. Due to this relatively limited number of exposures, data from this program were excluded from the training set. Instead, we utilized these observations for ASTERIS testing to evaluate the performance.

### *JWST NIRCam imaging data reduction*

We processed the NIRCam imaging using the JWST Science Calibration Pipeline (`jwst` v1.15.1) (*60*) and the Calibration Reference Data System (CRDS, (*87*)) mapping context file '`jwst_1318.pmap`'. We followed previous work by applying additional procedures to suppress the flicker noise (a form of correlated, low-frequency electrical noise, also known as the $1/f$ noise) (*61*), mitigate scattered light artifacts in the short-wavelength detectors, and for background estimation (*62*).

During pipeline execution at `jwst` Stage 1 (`calwebb_image` step), we set the jump detection parameters (`rejection_threshold`) to 3σ, with other parameters left as their default values. For `jwst` Stage 2, we applied default parameters. Based on the `jwst` stage-2 results, $1/f$ noise was subtracted at the amplifier level to suppress horizontal striping artifacts caused by electron drifting during readout. Astrometric correction was carried out using the TweakReg step, utilizing source coordinates detected within the same pointing and grouped by simultaneous exposures. The source coordinates were cross-matched with reference positions from the JADES NIRCam Photometry catalog Great Observatories Origins Deep Survey-South (GOODS-S) Deep v2.0 (*50*) for JOF images, and the Dark Energy Spectroscopic Instrument (DESI) Legacy Imaging Survey data release 10 (DR10) catalog (*63*) for the other datasets, to perform astrometric alignment.

Visual inspections were conducted to identify individual exposures exhibiting pronounced scattered light and periodic wave-like artifacts in the JOF field; the 27 affected exposures were excluded from our datasets. These artifacts were observed in F090W, F115W, F162M, F182M, F210M, F250M, and F335M bands. The remaining images were mosaicked into a unified target grid based on their WCS and resampled to a pixel scale of 0.04″ using the ResampleStep routine from the jwst Stage 3 pipeline. Outlier rejection when mosaicking was conducted with default parameters.

### *Subaru MOIRCS imaging data and reduction*

We used near-infrared imaging data in the Ks, BrG (Brackett-gamma, a near-infrared narrow band filter centered at 2.165 μm), and NB2083 (a near-infrared narrow band centered at 2.083 μm) filters taken by the MOIRCS (*6*, *47*, *48*, *88*) on the 8.2 m Subaru telescope. The observations were taken by program S17A-198S (*49*), S24A-091, and S24B-135. They were taken in May & June 2017, April & September 2024, and February 2025 under photometric conditions with seeing full width at half maximum (FWHM) ~ 0.7″. MOIRCS imaging data were first reduced using the Image Reduction and Analysis Facility (IRAF) (*89*) pipeline MCSRED (*90*). MCSRED was used to perform flat fielding, relative astrometry correction, a preliminary median sky subtraction, and distortion correction. Object masks were then constructed based on the pipeline output. After masking out detected objects, MCSRED was re-executed with additional polynomial sky subtraction. Based on the reduced individual exposure images, we used the SCAMP software (*91*) to perform absolute astrometry alignment to match the DESI Legacy Imaging Survey DR10 catalog, and we employed the SWARP software (*92*) to coadd the individual exposures. The data taken in 2017 and 2024 were used for model training, and the data taken in 2025 were used for testing. The results are shown in Fig. 3P-R.

## ASTERIS: model architecture and training strategy

### *Network architecture*

We developed two variants of the ASTERIS network—4-exposure (for wide-field surveys) and 8-exposure (for deep-field surveys)—chosen to match existing JWST observing programs. For both variants, we adopted a 3D U-Net architecture to enable multi-scale feature extraction and fusion, incorporating skip connections across both temporal and spatial dimensions (Fig. S1A) (*93*). For feature representation within CNNs, ASTERIS uses two Restormer-inspired (*37*) modules extended to the spatiotemporal domain: the 3D Multi-Deconvolved-Head-Transposed-Attention (3D-MDTA) block (Fig. S1D) and the 3D Gated-Deconvolved-Feed-Forward Network (3D-GDFN) block (Fig. S1E). The 3D-MDTA block introduced a self-attention mechanism that operates on convolution-extracted features, simultaneously leveraging the local connectivity and translation equivariance of the CNN while capturing long-range dependencies as in transformers.

This hybrid design was intended to circumvent the limited receptive field inherent to pure CNN, particularly for the structured, physically correlated background noise in astronomical imaging. Following 3D-MDTA, the 3D-GDFN block adaptively determines the relative importance of features, allowing downstream layers to focus on the most informative signals. This approach was chosen because in the astronomical background-limited regime, faint sources are dominated by bright extended sky background. Additionally, a dedicated 3D attention block was employed, at the network's output stage, to aggregate spatiotemporal features (Fig. S1F).

***Loss function***

ASTERIS was trained using a combination of average loss and frame loss. Average loss was defined as the mean squared error between the averaged inputs and targets. Frame loss was defined as the mean absolute error between individual input–target multi-exposure pairs. Formally, given a set of $M$-exposure inputs $\{X_i\}_{i=1}^{M}$, where $i$ denotes the exposure number, and corresponding targets $\{Y_i\}_{i=1}^{M}$, the average loss ($\mathcal{L}_{\mathrm{avg}}$) was computed as equation S1 and $\|.\|_2^2$ represents the sum of squares:

$$\mathcal{L}_{\mathrm{avg}} = \left\| \frac{1}{M}\sum_{i=1}^{M} X_i - \frac{1}{M}\sum_{i=1}^{M} Y_i \right\|_2^2 . \tag{S1}$$

This causes the network output to approximate the expectation of the underlying signal, enhancing source detection completeness by reducing the background noise fluctuations in the average image. The frame loss ($\mathcal{L}_{frame}$) was defined as equation S2 and $\| \|_1$ represents the sum of absolute values:

$$\mathcal{L}_{\mathrm{frame}} = \frac{1}{M}\sum_{i=1}^{M} \|X_i - Y_i\|_1 . \tag{S2}$$

Because the MAE loss corresponds to the conditional median estimation of the underlying pixel intensity distribution, $\mathcal{L}_{frame}$ provides robustness against outliers. This increases detection purity by suppressing false positive fluctuations in individual exposures (Fig. S7). The combined loss function for training ASTERIS was:

$$\mathcal{L}_{\mathrm{ASTERIS}} = \mathcal{L}_{\mathrm{avg}} + k\mathcal{L}_{\mathrm{frame}}, \tag{S3}$$

where the weighting factor $k$ affects both completeness and purity. We set $k = k_0 = 0.125$. This value was chosen to provide a balance between completeness and purity. To evaluate the effect of changing $k$, we conducted mock tests (as in Fig. 2) using models trained with $k = 0,\ 0.1k,\ k_0$, $10k_0$ and with frame loss only. We find that when $k$ is within an order of magnitude of $k_0$ ($0.1k_0 \leq k \leq 10k_0$), the overall performance (F-score) remained similar and outperforms either model trained with a single loss term (Fig. S7A-C). The model trained with frame loss only exhibited two distinct phases of sharp reduction during the evolution of loss functions, each followed by more stable plateaus (Fig. S7D-E). We interpret this behavior as due to the network's progression from learning low-frequency to high-frequency components, a phenomenon also seen in prior work (*94*).

### ***Data pre-processing for ASTERIS***

Following the data reduction of each observational dataset (see above), the multi-exposure images (after astrometric alignment) were combined along the temporal domain, to form a 3D data cube used as input for ASTERIS (Fig. 1B). We computed the standard deviation σ from the pixel distributions across all exposures after astrometric alignment, then a 3σ clipping was applied to the whole data cube in both spatial and temporal domains (outlier rejection). The clipping operation in the spatial domain separated the image stack into two components (Fig. S2): (a) a faint part, containing all pixel values with flux below 3σ significance, and (b) a bright part, composed of higher-value pixels that typically correspond to signals with intrinsically high S/N. Only the faint part was used for training and inference within ASTERIS. In contrast, the bright part was median co-added to remove residual cosmic rays and other artifacts. Before input into the network, the faint part was normalized by the *Z*-score:

$$Z_i = \frac{X_i - \mu}{\sigma}, \tag{S4}$$

where $\mu$ and $\sigma$ denote the median and standard deviation computed across the images. This normalization standardizes the dynamic range across different filters and programs, preserves the relative distribution of faint signals, and mitigates global brightness shifts that could bias network training. After denoising, the processed faint part was inversely *Z*-score normalized and recombined with the co-added bright part.

### ***Sigma-clipping threshold***

We investigated versions of ASTERIS trained with different sigma-clipping methods, which were evaluated by both experimental measurements and mock tests (Fig. S8). Without sigma-clipping, both completeness and purity are severely degraded (Fig. S8G-I). For the no sigma-clipping case, we investigated an alternative training strategy using a Poisson loss (`torch.nn.PoissonNLLLoss` in `PyTorch` (*95*, *96*)) instead of the average loss, which had been weighted. The Poisson negative log-likelihood loss, where the target count $g$ follows a Poisson distribution, is then defined as:

$$\mathcal{L}(\hat{g}, g) = \hat{g} - g\log(\hat{g} + \varepsilon) + \log(g!)\,, \tag{S5}$$

where $\hat{g}$ is the predicted mean count and $\varepsilon$ is a small constant added for numerical stability. Near the optimum $\hat{g} \approx g$, the gradient simplifies to:

$$\frac{\partial \mathcal{L}}{\partial \hat{g}} \approx \frac{\hat{g} - g}{\hat{g}}. \tag{S6}$$

Therefore, the same absolute error $(\hat{g} - g)$ at brighter pixels with larger $\hat{g}$ produces a smaller gradient. This causes `PoissonNLLLoss` to down-weight the contribution of bright pixels, balancing the influence of bright and faint sources. We found a similar improvement is achieved by weighting the loss functions with flux values (Fig. S8G-I). However, there the 3σ-clipping approach performs better in both completeness and purity. Clipping at 2σ instead led to larger numbers of outliers in the denoised images (Fig. S8C), primarily due to the high fraction of clipped outlier pixels in bright regions that are not denoised by the network. Clipping at 5σ provided slightly lower performance in purity compared to 3σ-clipping (Fig. S8F). Overall, we found that alternative strategies, such as Poisson-based losses or different clipping thresholds, didn't outperform the 3σ-clipping configuration across our tests.

***Training strategy***

JWST/NIRCam has two kinds of HgCdTe detectors with different characteristics (*7*, *97*). Therefore, the ASTERIS models were trained separately for the long-wavelength (filter central wavelength ≥ 2.5 μm) and short-wavelength (< 2.5 μm) NIRCam imaging data, yielding two different models for inference. After data pre-processing, the full imaging datasets collected from each JWST program were segmented into 120,000 patches (60,000 stack pairs), each comprising a data cube with 8 exposures of 128 × 128 pixels, for self-supervised training. During training, the dataset was divided into approximately 5,000 iterations, with a batch size of 12 stack pairs per iteration. The training was performed on four NVIDIA A100 GPUs with 40 GB GPU memory each. To maximize dataset diversity and avoid model overfitting, the eight input exposures for each patch were randomly selected from the full exposure set (≥ 16 exposures) before every training epoch. Each training pair was randomly augmented using multi-dimensional rotations and flips. Each training pair was also median-subtracted before training, to mitigate spatially varying local background effects and accelerate convergence. Both long- and short-wavelength models converged within 10 epochs using an initial learning rate of $1.5 \times 10^{-4}$. Training each model for 10 epochs required approximately 26 hours (batch size = 12).

For denoising (Figs. 3 to 6), ASTERIS automatically divides the input exposures into patches according to the specified patch size and the number of available GPUs. For an 8 × 1650 × 1650 pixel image data cube, we set the patch size to 8 × 800 × 800 with a batch size of four, running on the same hardware. The inference stage required approximately 18.1 seconds to produce a 1650 × 1650 pixel denoised output image. The resulting trained model of ASTERIS was found to be generally applicable across the adopted JWST NIRCam datasets from different programs (Fig. S12, Table S1), even those with different exposure times.

We applied ASTERIS models—trained separately on JWST and Subaru observations—to near-infrared (~ 2 μm) Subaru imaging data in the Ks band. While the JWST-trained model reduced noise, it introduced several false-positive detections (Fig. S14B). This is primarily attributed to the higher atmospheric scattering noise and turbulence-induced resolution degradation inherent in ground-based observations compared to space-based data. In contrast, the Subaru-trained model successfully recovered faint sources with few false positives, as verified by a deep co-addition image serving as the ground truth (Fig. S14C). Note that the Subaru training pipeline is same with the JWST discussed above and the training data spans multiple near-infrared filters (Ks, BrG, NB2083). To ensure reliability when adapting ASTERIS to different instruments, we recommend performing instrument-specific model training using corresponding observational data. The trained model for JWST NIRCam and Subaru MOIRCS is provided in (*86*).

## Quantitative mock tests

We benchmarked ASTERIS against previous denoising methods using mock tests on the JWST NIRCam imaging data from the GLIMPSE program (Fig. S4A). All co-addition baselines were implemented by averaging input exposures after outlier rejection (3σ clipping along the temporal domain following astrometric alignment), while the N2N baseline was implemented with the Restormer architecture (*37*). Mock sources were generated using the empirical PSF generated from the `STPSF` software (*64*). The FWHM of the empirical PSF is approximately 0.036″, and the radius enclosing 80% of the encircled energy (EE80) is approximately 0.136″ (Fig. S4C). The apparent magnitudes of the mock sources are distributed between 27.5 and 31.5 mag, following a third-order power-law distribution, consistent with the observed source counts in the real data (Fig. S4B). Some relatively bright pixels in the injected sources (≤ 30 mag) were clipped during the 3σ-clipping prior to ASTERIS and then recombined after denoising (Fig. 1B). The parameters used for the mock tests are summarized in Table S2.

The performance is quantified using the completeness and purity. Completeness is defined as the ratio of true-positive detections of injected sources to the total number of mock sources. Purity is defined as the ratio of false-positive detections to the total number of positive detections. We also calculate the F-score (Fig. S5K), which considers both completeness and purity (*65,66*):

$$\text{F-score} = 2 \times \frac{\text{Completeness} \times \text{Purity}}{\text{Completeness} + \text{Purity}}. \quad (S7)$$

To identify sources in the image we used `Source Extractor` v2.25.0 (*38*), from which we determined the number of true-positive and false-positive detections. We set the source detection threshold to at least eight connecting pixels above 0.5σ significance. This low significance threshold was set to facilitate as many source detections as possible for a better evaluation of the completeness. Cross-matching was conducted against sources identified in the ground-truth injected sources, using a maximum separation threshold of 0.04″ (1 pixel). Mock source detections matched to ground-truth sources were classified as true positives, while unmatched detections were designated as false positives. The PSFs produced by each method in Fig. 2I were extracted using the `PSFEx` software (*67*). To evaluate the photometric accuracy of ASTERIS, we performed forced photometry above the ground-truth source locations on co-addition and ASTERIS results via a 0.14″ circular aperture to measure the magnitude difference (Fig. 2J).

## High-redshift Galaxies Identification and UV Luminosity Function

### *Adopted Cosmology and Conventions*

We adopt a Λ cold dark matter (ΛCDM) cosmology with Hubble constant $H_0 = 70\ \text{km s}^{-1}\text{Mpc}^{-1}$, matter density parameter $\Omega_m = 0.3$, and dark energy density parameter $\Omega_\Lambda = 0.7$. All magnitudes used the AB system, so the apparent magnitude $m_{\text{AB}} = -2.5\log(f_\nu/\text{nJy}) + 31.4$, where $f_\nu$ is the flux density in nanojansky.

### *Source Extraction and Photometry*

We searched for high-redshift galaxy candidates in the JOF deep imaging field observed with JWST. To perform source extraction, we first constructed a detection image by directly combining the ASTERIS-denoised mosaics in three NIRCam long wavelength wide-band filters: F277W, F356W, and F444W. Then, we ran `Source Extractor` on the detection image with a threshold of 1.3σ significance and a minimum deblending contrast parameter of 0.0001. Forced aperture photometry was conducted with a 0.1″ radius circular aperture. An aperture correction was applied to account for flux outside the aperture, using the PSF profile and the expected flux ratio between a circular aperture and Kron photometry (*68*). Photometric uncertainties were estimated using random aperture sampling (0.1″ radius, 10,000 times) on clear, source-free regions, which were masked by the segmentation image generated by `Source Extractor`. These methods were chosen to match the previous works we use for comparison (*52*, *53*).

***Sample Selection***

High-redshift galaxies were selected using the Lyman-break technique (*54*), whereby intervening neutral hydrogen absorbs ultraviolet light blueward of the Lyman-$\alpha$ emission line (1215.6 Å in the rest frame). Due to cosmic expansion, the Lyman-α break is redshifted into the near-infrared wavelength for $z \gtrsim 9$ objects, causing them to appear bright in longer-wavelength JWST NIRCam filters but invisible (dropout) in shorter-wavelength filters. We adapted the selection criteria from previous studies (*53*) to select high-redshift galaxy candidates. All sources from the photometric catalog were initially selected using color criteria, followed by a preliminary spectral energy distribution (SED) fitting using the `EAZY` software (*69*) and a visual inspection.

We focused on high-redshift galaxies at $z \sim 9$ to 22.5, the range for which the Lyman-$\alpha$ break is redshifted into the wavelengths covered by three NIRCam wide filters, F115W ($z \sim 9$ to 12), F150W ($z \sim 12$ to 16), and F200W ($z \sim 16$ to 22.5). To identify dropouts in each filter, a different set of color selection criteria was employed:

1. Dropout color criterion: The AB magnitude difference between the dropout band and the adjacent redder wide band must be > 1.3 mag. Specifically, F115W- F150W ≥ 1.3 mag for F115W dropout sources, F150W-F200W ≥ 1.3 mag for F150W dropout sources, and F200W-F277W ≥ 1.3 mag for F200W dropout sources.

2. Robust detection in the rest-frame UV and optical bands: The source must be robustly detected in filters that correspond to rest-frame UV and optical wavelengths. Specifically, S/N > 2 was required in all wide-band filters redder than the dropout band, and S/N > 5 was required in more than half of these filters, including the F410M filter.

3. Non-detection in bluer bands: The source must not be detected in any of the wide-band filters that are bluer than the dropout band.

4. Rest-frame UV color constraint: The source must not exhibit extremely red colors in the filters probing the rest-frame UV. Specifically, F150W - F277W < 1.0 mag for F115W dropout sources, F200W - F356W < 1.0 mag for F150W dropout sources, and F277W - F444W < 1.0 mag for F200W dropout sources.

5. Dropout color dominance requirement: The color break across the dropout must be substantially stronger than the rest-frame UV color. Specifically, F115W-F150W ≥ F150W-F277W + 1.3 mag for F115W dropout sources, F150W-F200W ≥ F200W-F356W + 1.3 mag for F150W dropout sources, and F200W-F277W ≥ F277W-F444W + 1.3 mag for F200W dropout sources.

This color selection was carried out using the photometry with radius 0.1″ circular apertures, which is sufficient to enclose the compact sizes of most high-redshift galaxies (*98–102*). For photometry results with S/N ≤ 1, we set the flux to the 1σ upper limit, to avoid infinite magnitude values in the comparisons.

We then estimated the photometric redshifts for all sources that satisfy the color selection criteria via SED fitting. We adopted an existing template suite (*70*), incorporating models with strong emission lines and a range of UV continua, and an intergalactic medium attenuation model (*103*). All 14 bands were used for EAZY SED fitting; the potential redshift range was set to [0.01, 22.50] with a step size of 0.01. No magnitude priors were applied during the fitting process; a minimum uncertainty of 5% was imposed on the photometry, and negative flux values were permitted. We required that candidate sources have redshift probability of > 50% at $z \geq 8$ for F115W dropouts, at $z \geq 11$ for F150W dropouts, and at $z \geq 15$ for F200W dropouts. We adopted the redshift solution corresponding to the minimum $\chi^2$ from the SED model fitting, as the initial photometric redshift estimate. All potential high-redshift galaxy candidates that satisfy both the color selection and photometric redshift criteria were visually inspected to remove those contaminated by imaging artifacts and nearby bright sources. The source ID, coordinates, and dropout filter of final selected high-redshift galaxy candidates are provided in Table S3 and Data S1. The SED fitting results, photometric measurements, and image cutouts of some examples are shown in Fig. 4-5, S16.

### *UV Luminosity Measurement*

We inferred the physical properties of the galaxies in the color-selected sample by fitting the NIRCam photometry with the Bayesian Analysis of Galaxy SEDs (`BEAGLE`) software (*71*) using the initial redshift solutions produced by `EAZY`. This stage followed the modeling methods and parameters described in previous work (*53*). For the set of SED models used in `BEAGLE`, we adopted an initial mass function (*104*) from 0.1 to 300 solar masses ($M_\odot$). We used the Small Magellanic Cloud dust attenuation curve (*105*) with V-band optical depth $\tau_V$ in the range $-3 \leq \log_{10}\tau_V \leq 0.7$, then assumed the same intergalactic attenuation model (*103*) as above. For the stellar properties, we used log-uniform priors on maximum stellar age ($t_{max}$) from $7.35 \leq \log_{10}(t_{max}/\mathrm{yr}) \leq \log_{10}(t_{univ}(z_{phot})/\mathrm{yr})$ where $t_{univ}$ is the age of the Universe at redshift $z_{phot}$, on stellar mass ($M_*$) from $5 \leq \log_{10}(M_*/M_\odot) \leq 12$, on ionization parameters ($U$) in the range of $-4 \leq \log_{10}(U) \leq -1$, and on metallicity ($Z$) spanning $-2.2 \leq \log_{10}(Z/Z_\odot) \leq 0.3$ where $Z_\odot$ is the solar metallicity. For the star formation history (SFH), we assumed a two-component model, a constant model and a delayed exponential model. We set the delayed exponential $e$-folding time ($\tau$) as $6 \leq \log_{10}(\tau/\mathrm{yr}) \leq 10.5$, the duration of recent constant SFH component ($t_{recent}$) as $6 \leq \log_{10}(t_{recent}/\mathrm{yr}) \leq 7.3$, and the specific star formation rate of recent constant SFH component ($\mathrm{sSFR}_{recent}$) as $-14 \leq \log_{10}(\mathrm{sSFR}_{recent}/\mathrm{yr}^{-1}) \leq -6$.

Following previous work (*53*), we restricted the redshift limits for F115W, F150W and F200W to 8, 11 and 15 respectively. From the best-fitting SEDs, we measured $z_{phot}$ as the 50 percentiles from the posterior probability distribution, with uncertainties estimated as the 16 to 84 percentiles. To determine the UV luminosity of each galaxy candidate, we require the quantities $z_{phot}$ and $M_{UV}$, with their uncertainties. We calculated the UV luminosity by integrating the best-fitting model spectra over the rest-frame wavelengths $\lambda_{rest}$ =1450 to 1550 Å, then used the best-fitting $z_{phot}$ to convert this to $M_{UV}$.

### *Completeness*

To assess the incompleteness resulting from the source detection and galaxy selection, we performed a mock test of source injection and recovery using the same methods as above. We injected mock sources with different fluxes and shapes into the real JOF images, then used the same techniques for detecting these sources, measured their photometry, and selected samples as we did with the real data. We followed previous methods (*53*) to inject artificial sources onto the real JOF mosaic images. The injected sources had a power-law spectrum $f_\lambda \propto \lambda^\beta$, where $\lambda$ denotes the wavelength, $\beta$ is the rest-frame UV slope. The flux was derived from a uniform distribution of absolute UV magnitudes between $M_{UV}$ = -24 to -15 and a uniform distribution of redshifts in the range $z = 7$ to 25, which encompass the flux and redshift range of the real sample. The rest-frame UV slope $\beta$ was determined from an empirical $\beta$-$M_{\rm UV}$ relation (*106*). We applied the same intergalactic medium (IGM) attenuation model (*103*) to the SED and measured the model photometry by integrating the SED over the transmission curve of each filter band used in the JOF data. The mock galaxy sizes were determined from an empirical size-UV luminosity relation (*114*). The injected source shapes had a Sérsic profile (*107*) with Sérsic index $s$ randomly drawn from a one-sided truncated Gaussian distribution with mean $\mu_s = 1$, standard deviation $\sigma_s = 1$, and truncated interval from 0.5 to $\infty$. The axis ratio was randomly drawn from a truncated Gaussian distribution with mean $\mu_q = 0.8$, standard deviation $\sigma_q = 0.4$, and a truncated interval from 0 to 1. The position angles were randomly determined from a uniform distribution within a range from -90° to +90°.

To estimate the source detection completeness, we produced a mock source catalog with $2 \times 10^5$ sources using the same sampling method as described above. Then, for each mock galaxy, we made a cutout image centered at a random position on the real JOF mosaics with a size of 8″ × 8″ (200 × 200 pixels). We then used the `GALSIM` software (*108*) to inject these mock galaxies onto the cutout images. We ran the same source detection and photometric measurement processes as we used for the real images. To estimate the selection completeness, we produced a larger mock source catalog with $1.26 \times 10^6$ sources and repeated the photometric selection processes. The total completeness function $C\,(M_{\rm UV}, z)$ varies with absolute UV magnitude and redshift (Fig. S17), calculated as the fraction of recovered mock sources following the same methods as used for the real sources. The results have a maximum completeness of about 78 percent, due to overlapping objects.

***UV Luminosity Function***

We used the effective volume method (*72*) to calculate the binned luminosity functions (LFs). In this approach, the completeness factor is combined with the effective survey volume ($V_{\rm eff}$), to correct the derived luminosity functions. The luminosity function $\Phi(M_{\rm UV})$ in each absolute UV magnitude bin $\mathrm{d}M_{\rm UV}$ is calculated by:

$$\Phi(M_{\rm UV})\mathrm{d}M_{\rm UV} = \sum_j \frac{1}{V_{\rm eff}\left(M_{{\rm UV},j}\right)}, \qquad \text{(S8)}$$

where the sum extends over all sources, indexed by $j$, with absolute UV magnitude $M_{{\rm UV},j}$ falling within the bin. The effective volume $V_{\rm eff}$ as a function of absolute UV magnitude is given by:

$$V_{\rm eff}(M_{\rm UV}) = \int_{z_{\rm min}}^{z_{\rm max}} \frac{\mathrm{d}V_{\rm com}}{\mathrm{d}z\mathrm{d}\Omega} C(M_{\rm UV}, z)\Omega(z)\mathrm{d}z\,, \qquad \text{(S9)}$$

where $z_{\rm min}$ and $z_{\rm max}$ are the lower and upper redshifts, respectively, $\mathrm{d}V_{\rm com}$ is the comoving volume element per unit area $\mathrm{d}\Omega$ at a redshift $z$, and $\Omega(z)$ is the survey area at redshift $z$. For our sample, the corresponding survey area is 9.08 arcmin$^2$, which is calculated by counting all pixels with valid exposures ($2.043\times10^7$ pixels in total, each pixel represents for 0.04″ × 0.04″) in JOF. Due to the substantial uncertainty of the photometric redshifts $z_{\rm phot}$ and absolute UV magnitudes $M_{\rm UV}$, we take them into account in calculating the luminosity function. We ran Monte Carlo (MC) sampling and assumed each quantity was sampled from a Gaussian distribution with its uncertainty as the standard deviation. The final LF uncertainty in each magnitude bin was propagated from the Poisson error and the MC sampling. For the case where $\Phi$ is smaller than its uncertainty, we instead used the 2σ upper limits, defined as the inverse survival function of the Gamma distribution.

The final LFs of F115W and F150W dropouts are shown in Fig. 6A-B. We did not derive a luminosity function for the F200W dropouts, as the sample size was insufficient to provide statistical constraints.-The mean completeness in the faintest bin is about 10%, which is $M_{UV}$ = -16.1 for F115W dropouts and $M_{UV}$ = -16.5 for F150W dropouts.

The selected samples could be affected by fluctuations due to large-scale structure (cosmic variance). We estimated a cosmic variance of ~ 20% for our sample using the `Cosmic Variance Calculator` v1.03 software (*109*), which is numerically comparable to the relative Poisson uncertainty (>20% for a sample of less than 25 sources in the bin). This does not affect our conclusions.

We fitted the binned luminosity functions with a model Schechter function (*73*),

$$\Phi(M_{\mathrm{UV}}) = 0.4\ln(10)\,\Phi^{*}10^{-0.4(M_{\mathrm{UV}}-M_{\mathrm{UV}}^{*})(\alpha_s+1)} \times \exp\left[-10^{-0.4(M_{\mathrm{UV}}-M_{\mathrm{UV}}^{*})}\right], \quad \text{(S10)}$$

where $\Phi^*$ is the overall normalization, $M_{\mathrm{UV}}^*$ is the characteristic UV luminosity, and $\alpha_s$ is the faint end slope. Because our sample is dominated by galaxy candidates at the faint end ($M_{\mathrm{UV}} > -18$), we cannot constrain the $M_{\mathrm{UV}}^*$ parameter, which is typically $M_{\mathrm{UV}}^* < -20$ (*53*). Therefore, we fixed the characteristic luminosity $M_{\mathrm{UV}}^*$ to the value $M_{\mathrm{UV}}^*$ = -20.32 found by the JADES collaboration (*53*). Model fitting therefore had two free parameters: the overall normalization and the faint end slope. For the F115W dropouts, we found $\Phi^* = 7.09 \times 10^{-5}$ Mpc$^{-3}$ mag$^{-1}$ and $\alpha_s$ = - 2.45 ± 0.03. For the F150W dropouts, we found $\Phi^* = 3.82 \times 10^{-5}$ Mpc$^{-3}$ mag$^{-1}$ and $\alpha_s$ = - 2.28 ± 0.02.

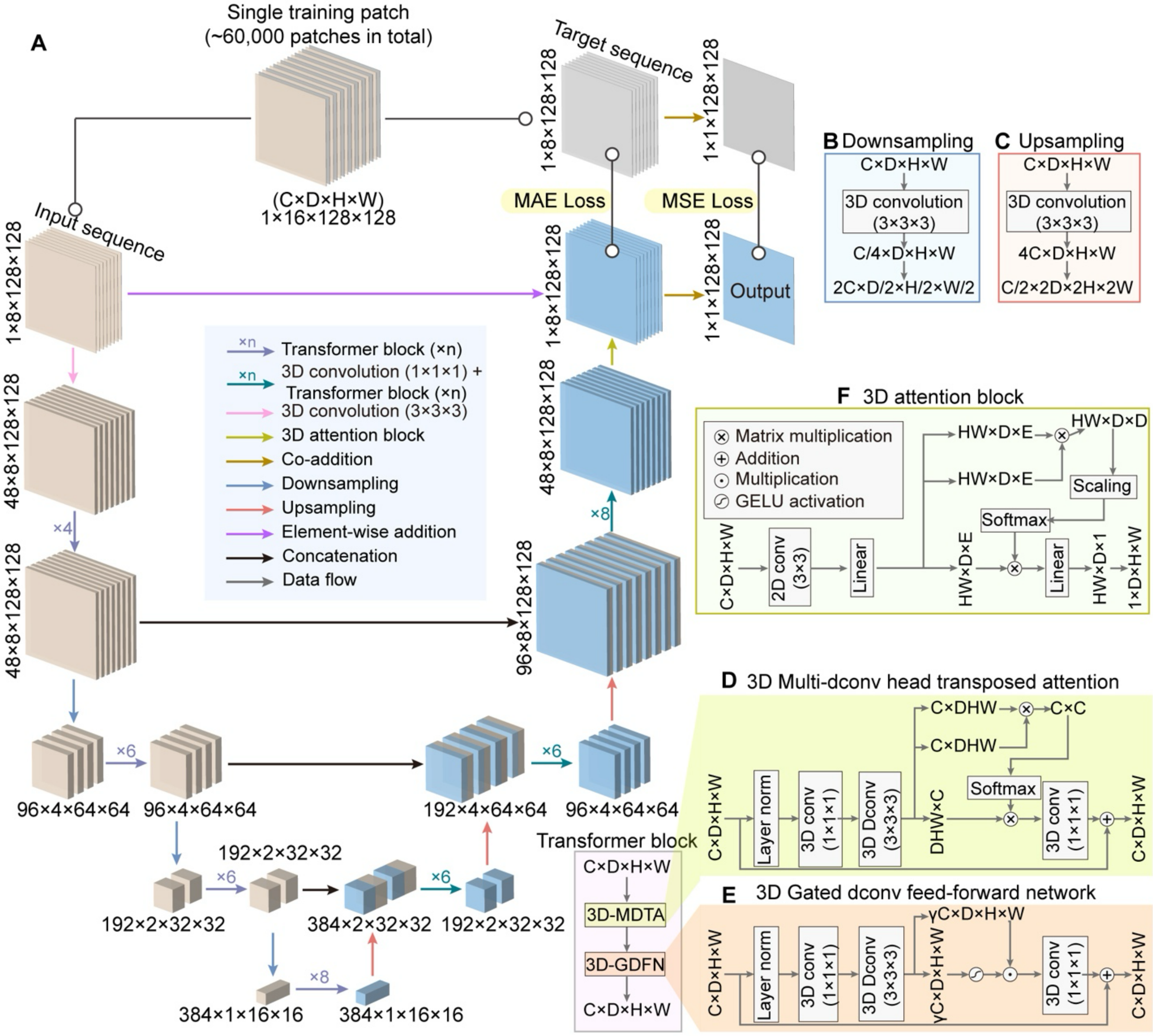

**Fig. S1. Network architecture of ASTERIS.** ASTERIS is designed for hierarchical spatiotemporal feature extraction and image denoising based on Restormer (*37*). (**A** to **C**) The architecture follows a 3D U-Net architecture consisting of an encoder (left column of panel A) and a decoder (right column of panel A). Feature map dimensions are labeled as C × D × H × W (Channels×Depth×Height×Width). The encoder progressively downsamples the input sequence (1×8×128×128) to capture high-level features, while the decoder symmetrically upsamples these features and generate the final denoised output. Multipliers (e.g., ×4, ×6, ×8) indicates the repetition of Transformer blocks or hybrid 3D convolution-Transformer units as defined in the legend. Visual indicators are used to define the data flow and operations: black arrows denote general data flow; purple arrows represent 3D convolutions (3×3×3); blue arrows and red arrows signify downsampling (panel B) and upsampling (panel C) processes, respectively; black arrows indicate feature concatenation (skip connections) that bridge the encoder and decoder stages; and brown arrows represent co-addition (average). The network integrates three primary functional modules: (**D**) the 3D Multi-dconv head Transposed Attention (3D-MDTA) block, where the Layer norm represents layer normalization and Softmax denotes a normalization function; (**E**) the 3D

norm represents layer normalization and Softmax denotes a normalization function; (**E**) the 3D Gated dconv Feed-Forward Network (3D-GDFN) block, where the $\gamma$ represents expansion ratio of the feature channels; and (**F**) a 3D attention block for global spatiotemporal aggregation, where the linear block represents a linear projection, scaling block denotes a normalization step in the scaled dot-product attention mechanism, E represents the feature dimension, and GELU means Gaussian Error Linear Unit. Training is supervised by MAE and MSE loss functions.

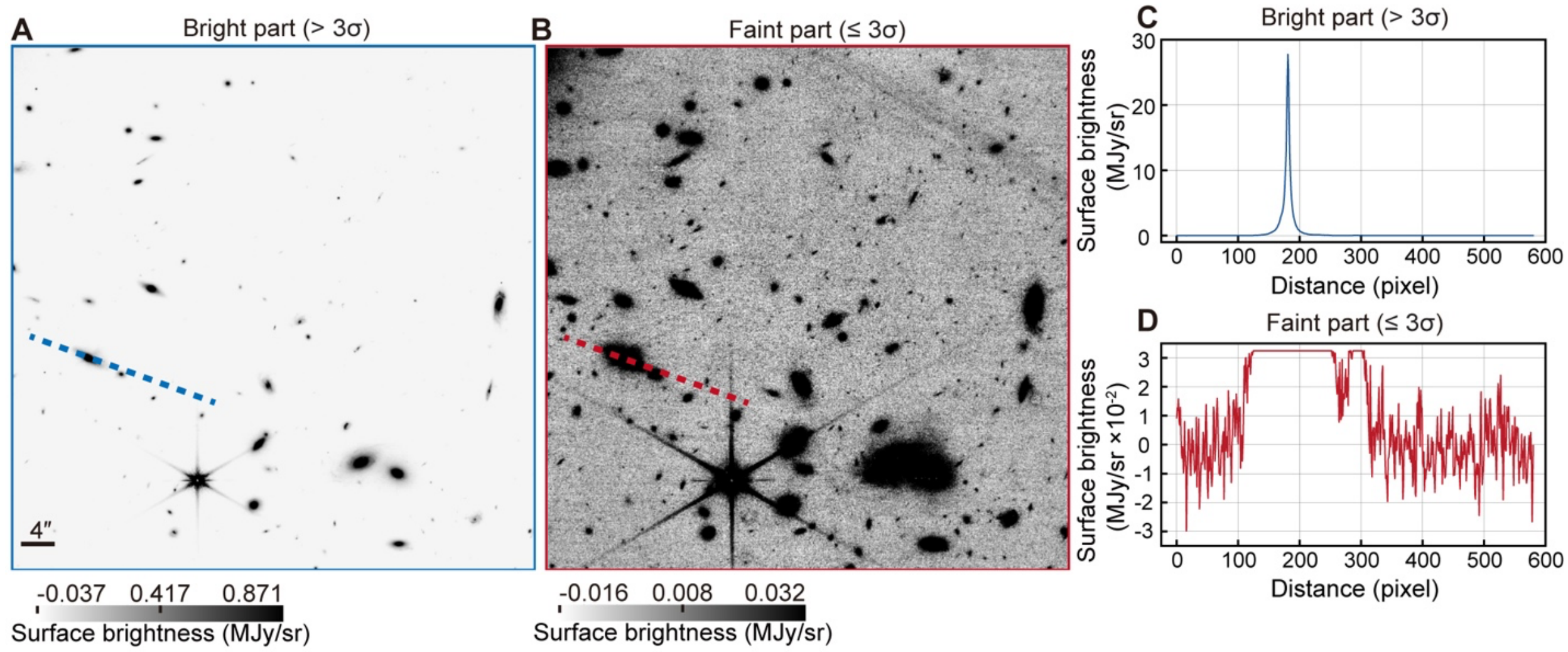

**Figure. S2. Illustration of 3σ-clipped images before running ASTERIS. (A)** Bright part of an example JWST GLIMPSE program NIRCam F150W image extracted by the 3σ clipping. Scale bar, 4 arcseconds. **(B)** Faint part of the same NIRCam image extracted by the 3σ clipping. **(C)** Trace plot along the dashed line in panel A. **(D)** Trace plot along the dashed line in panel B, which is at the same location as in panel A.

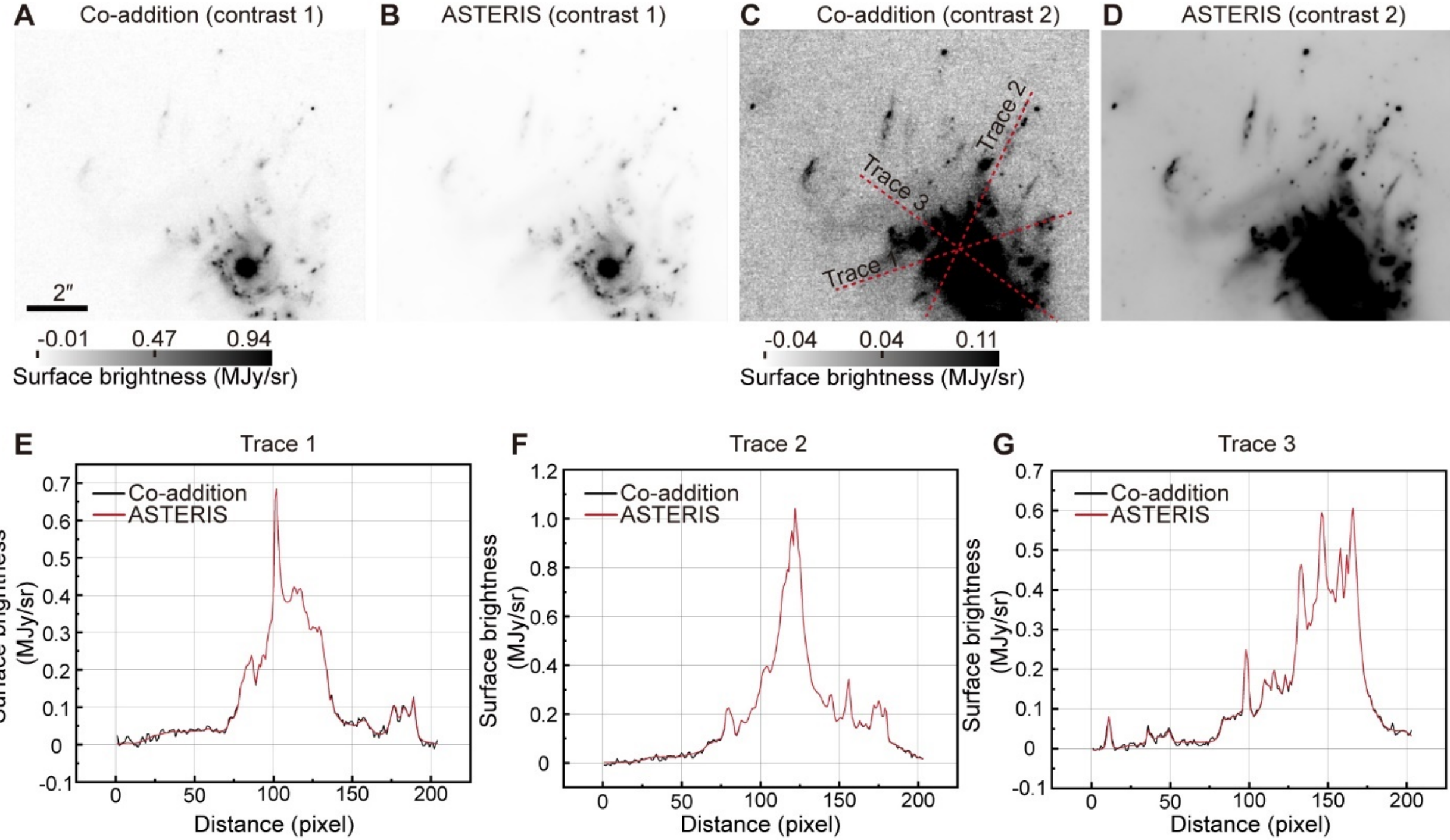

**Figure. S3. Evaluation of the transition between bright and faint regions in ASTERIS.** The image is from the JWST Mega Science Survey (Program ID 4111). **(A–D)** NIRCam F115W imaging from co-addition and ASTERIS, shown at two contrasts with different colorbars to highlight the bright and faint regions, respectively. Scale bar, 2 arcseconds. **(E-G)** Trace plots along the red dashed lines in panel B.

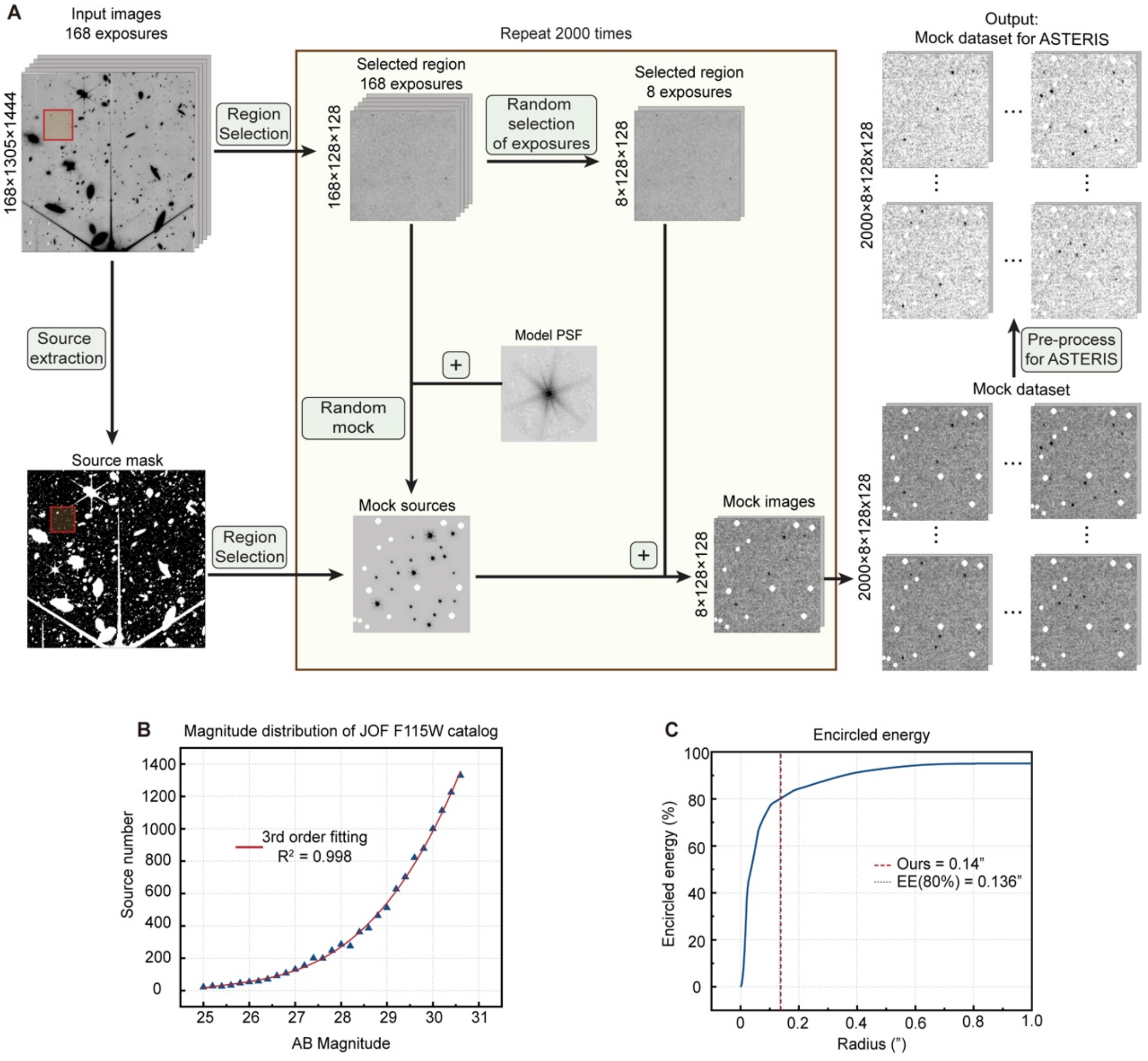

**Fig. S4. Mock data generation for quantitative evaluation. (A)** Schematic flowchart of quantitative mock data generation. Utilizing the JWST NIRCam imaging data from the GLIMPSE program, the mock dataset generation begins with 168 independent exposures (1305 × 1444 pixels each). A source mask is generated from the input images using `Source Extractor`, and a source-free region of 128 × 128 pixels is selected (red boxes). The selected region is simultaneously cutout from both input images and source mask. Mock sources are randomly generated using the model PSF from `STPSF`, and then added to a random set of eight background exposures chosen from the 168 input images. This operation is repeated 2000 times (yellow shaded box) to produce the mock dataset, which subsequently follows the same pre-processing pipeline, including sigma-clipping and Z-score normalization. **(B)** The apparent magnitude distribution (blue triangles) of detected sources in F115W of the JOF NIRCam data. A third-order polynomial fitted to the distribution (red line) has a coefficient of determination ($R^2$) of 0.998. In mock tests, the apparent magnitudes of mock sources follow the same power-law distribution. **(C)** The encircled energy (EE) profile of the empirical PSF in mock tests. The radius enclosing 80% of the encircled energy is 0.136″ (vertical dotted red line). A circular aperture with 0.14″ radius is employed for photometry, corresponding to approximately 83% encircled energy.

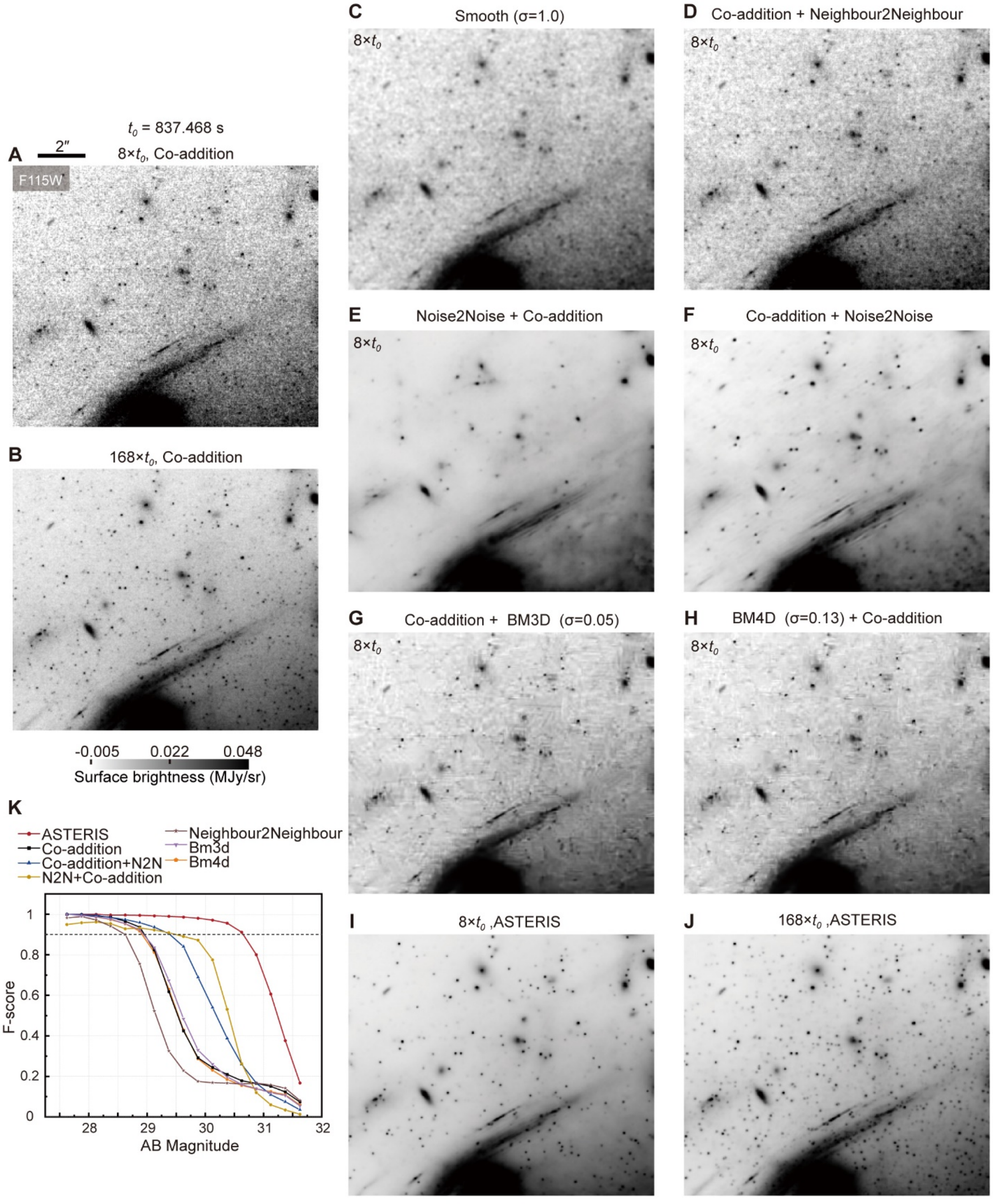

**Fig. S5. Comparison between ASTERIS and previous denoising methods.** Output of each denoising method applied to a crowded field near a bright source, taken from the JWST GLIMPSE program NIRCam F115W imaging. $t_0$ is the exposure time per exposure (837.5 s). M× indicates that M exposures were used as input for each method. Scale bar, 2 arcseconds. **(A)** Co-addition result using 8×$t_0$ exposures. **(B)** Deep co-addition results from 168×$t_0$ exposures, serving as the ground truth. **(C)** Smoothed version of panel A, using a Gaussian kernel with σ = 1.0 pixels. **(D)** Neighbor2Neighbor (*27*) applied to the image in panel A. **(E)** Noise2Noise (*30*) applied independently to each of the $8 \times t_0$ exposures, followed by co-addition after denoising. **(F)** Noise2Noise applied directly to the image in panel A. **(G)** BM3D (*35*) applied to the image in panel A, with σ = 0.05. **(H)** BM4D (*36*) applied to the full $8 \times t_0$ exposure stack, followed by co-addition after denoising; with σ = 0.13. **(I)** ASTERIS result using $8 \times t_0$ exposures. **(J)** ASTERIS result using $168 \times t_0$ exposures. For this, groups of 21 exposures were co-added into a single image, producing 8 exposures in total, used as ASTERIS input. **(K)** Comparison of F-score as a function of apparent magnitude for each method (colored lines, see legend). The horizontal dashed black line marks an F-score of 0.9.

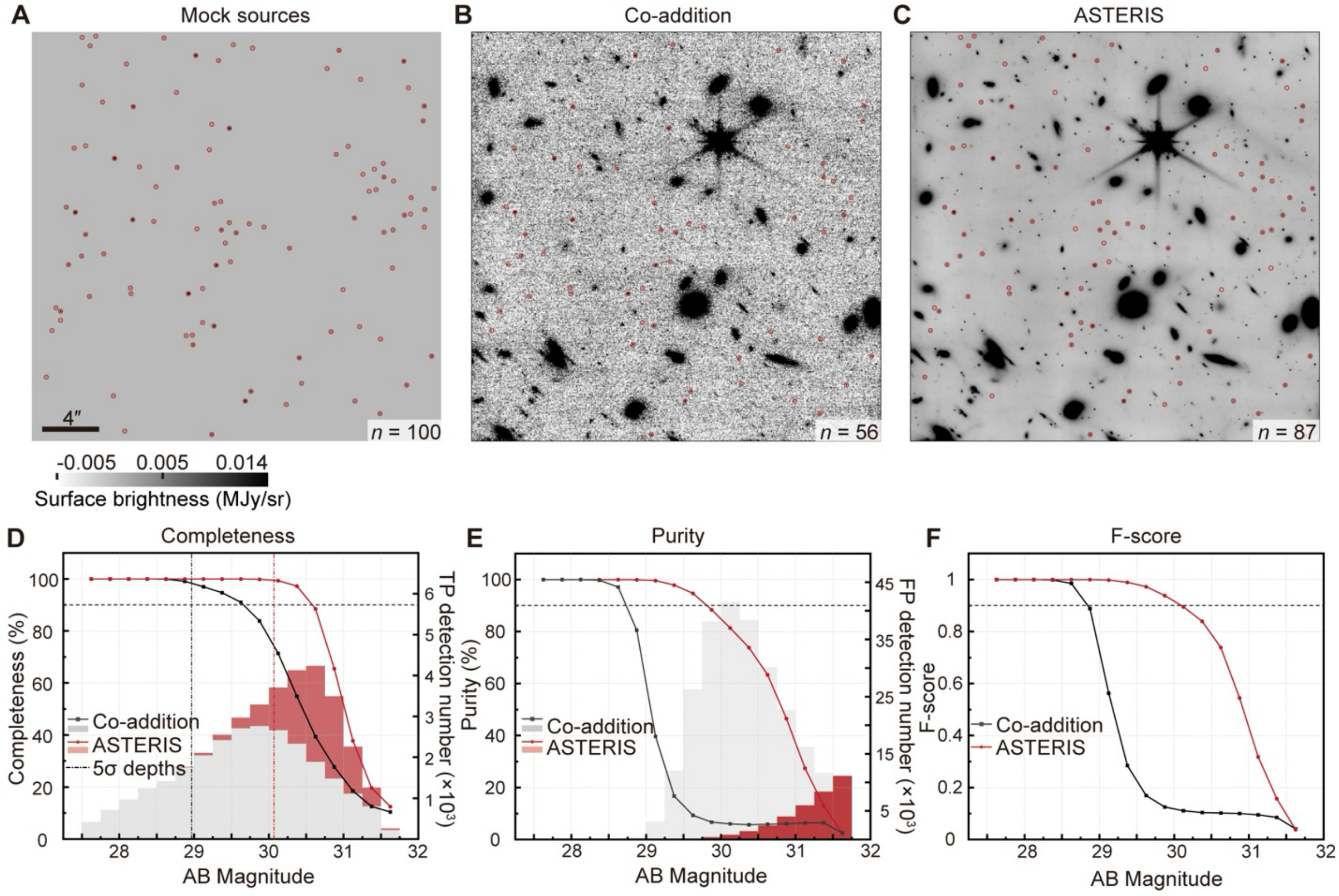

**Fig. S6. Mock tests of ASTERIS on complex backgrounds.** An example of mock source injection into the JWST GLIMPSE program NIRCam F115W imaging. Similar to Fig. 2, but for a more complicated background image. **(A)** Ground-truth image of the injected sources. **(B)** Output image from standard co-addition (8 exposures averaged). **(C)** Output image from ASTERIS (8 exposures). The labeled values *n* indicates the number of detected sources in each image (red circles), using `Source Extractor` with identical parameters. Scale bar, 4 arcseconds. **(D-F)** Same as Fig 2F-H, but applied to this field.

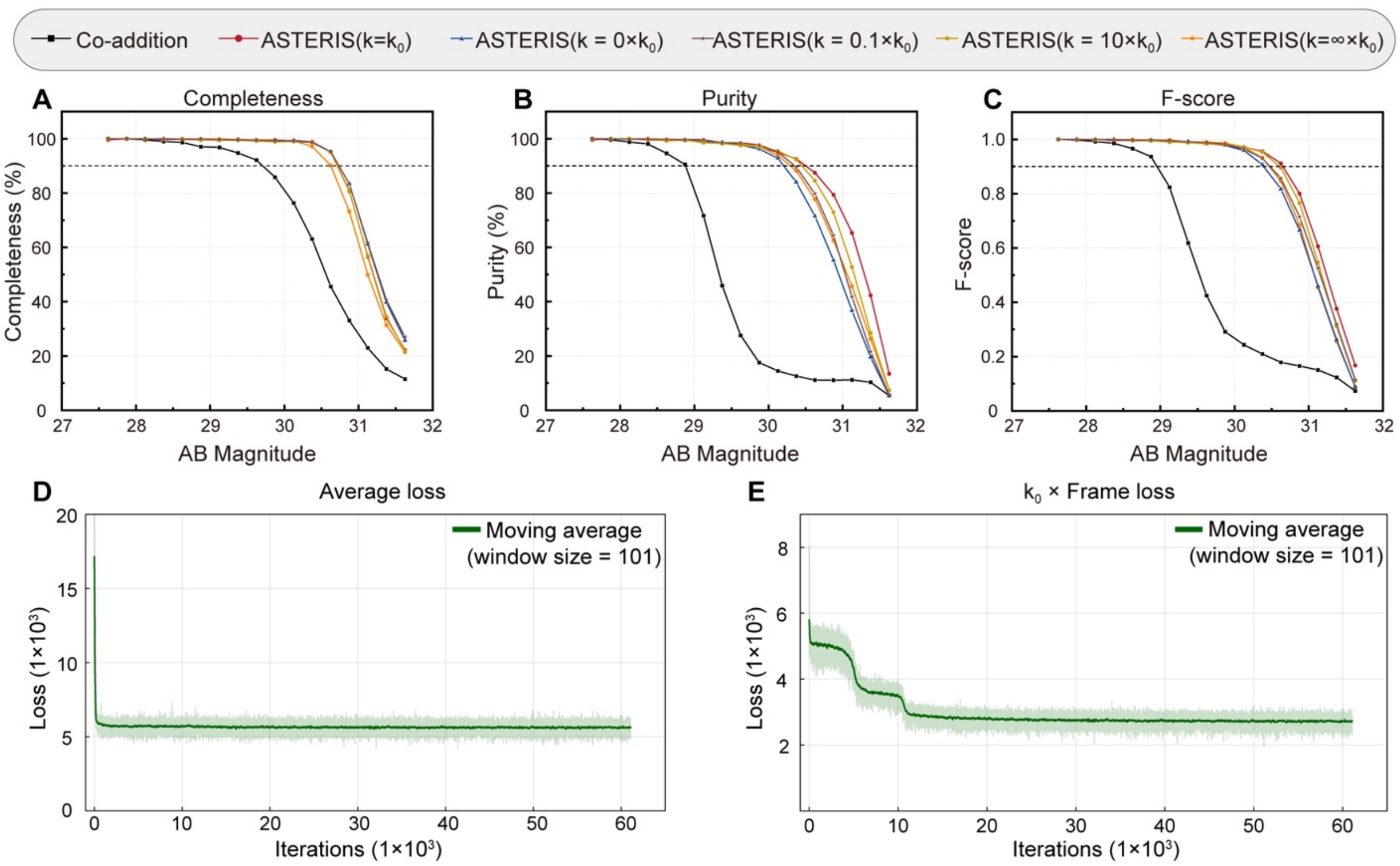

**Fig. S7. Ablation study of the ASTERIS loss function.** ASTERIS was trained using a combination of the average loss (equation S1) and frame loss (equation S2). Loss of ASTERIS is the combination of both average loss and frame loss with an empirical weight factor $k$. **(A)** Detection completeness from the mock tests using models trained with $k = 0$, $0.1k_0$, $k_0$, $10k_0$, and $\infty \times k_0$ (with the frame loss only) (colored lines, see legend), $k_0 = 0.125$ is an empirical value; the 90% completeness threshold is indicated by a black dashed line. **(B)** Same as panel A but for detection purity; the 90% completeness threshold is indicated by a black dashed line. **(C)** Same as panel A but for F-scores. The 0.9 F-score is shown as a black dashed line. **(D)** Evolution of the average loss term and, (**E**) the frame loss term, both as functions of training iterations with $k = k_0$.

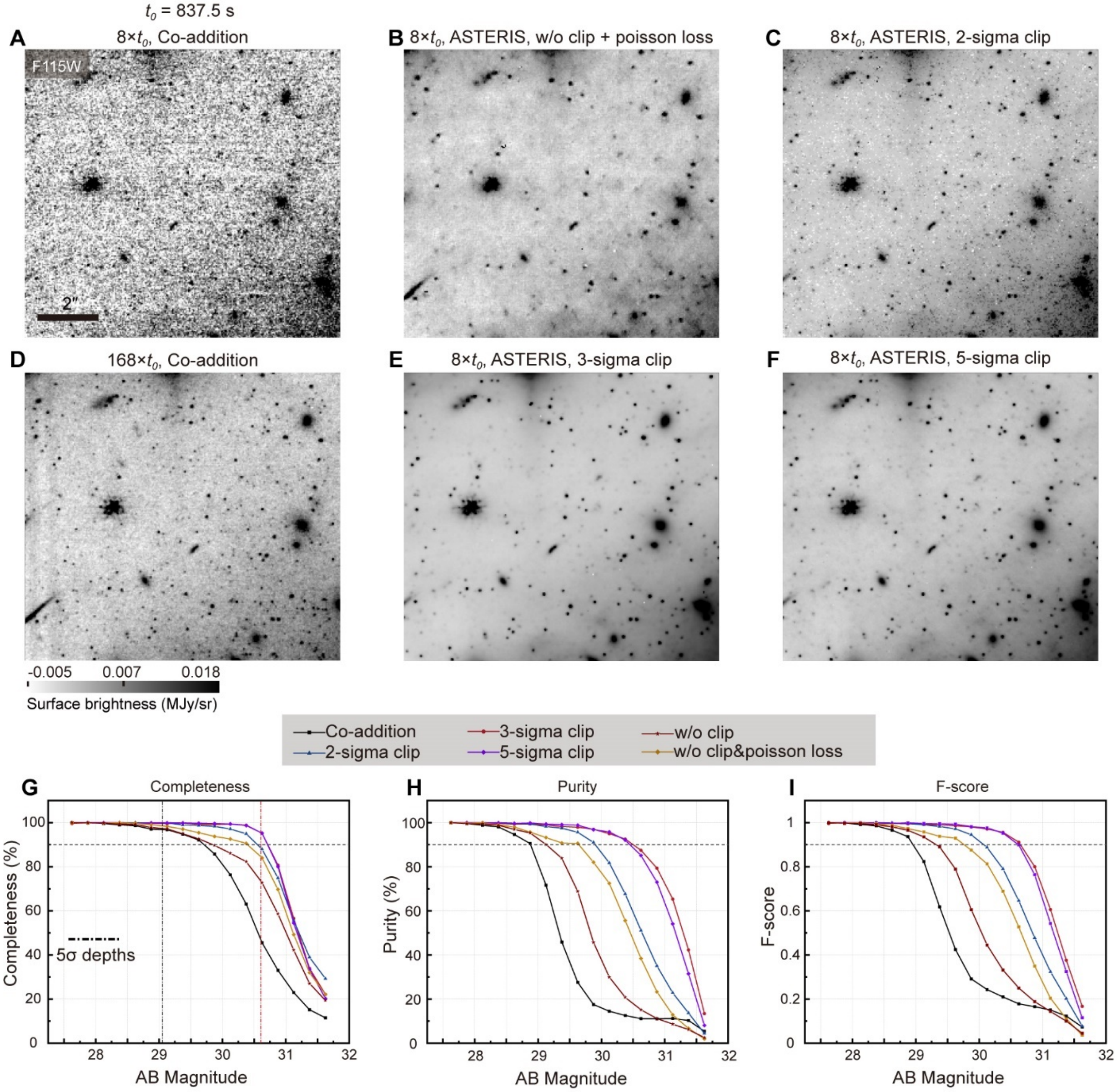

**Fig. S8. Comparison of sigma-clipping methods used in ASTERIS. (A to F)** Images using different sigma-clipping methods on a crowded field from the JWST GLIMPSE program NIRCam F115W imaging. **(A)** Standard co-addition using $8\times t_0$ exposures. **(B)** ASTERIS trained with Poisson loss without sigma-clipping preprocessing. The Poisson loss is induced to balance the influence on the network between bright sources and faint sources. **(C)** ASTERIS with 2σ-clipping preprocessing. **(D)** Deep co-addition result using $168\times t_0$ exposures, serving as the ground truth. **(E)** ASTERIS with 3σ-clipping preprocessing. **(F)** ASTERIS with 5σ-clipping preprocessing. **(G to I)** Quantitative comparison of the different sigma-clipping methods from the mock tests in Fig. 2. **(G)** Completeness comparison. The 90% completeness level is plotted by a black dashed line. **(H)** Purity comparison. The 90% purity level is plotted by a black dashed line. **(I)** F-score comparison. The F-score of 0.9 is shown as a black dashed line. Scale bar, 2 arcseconds.

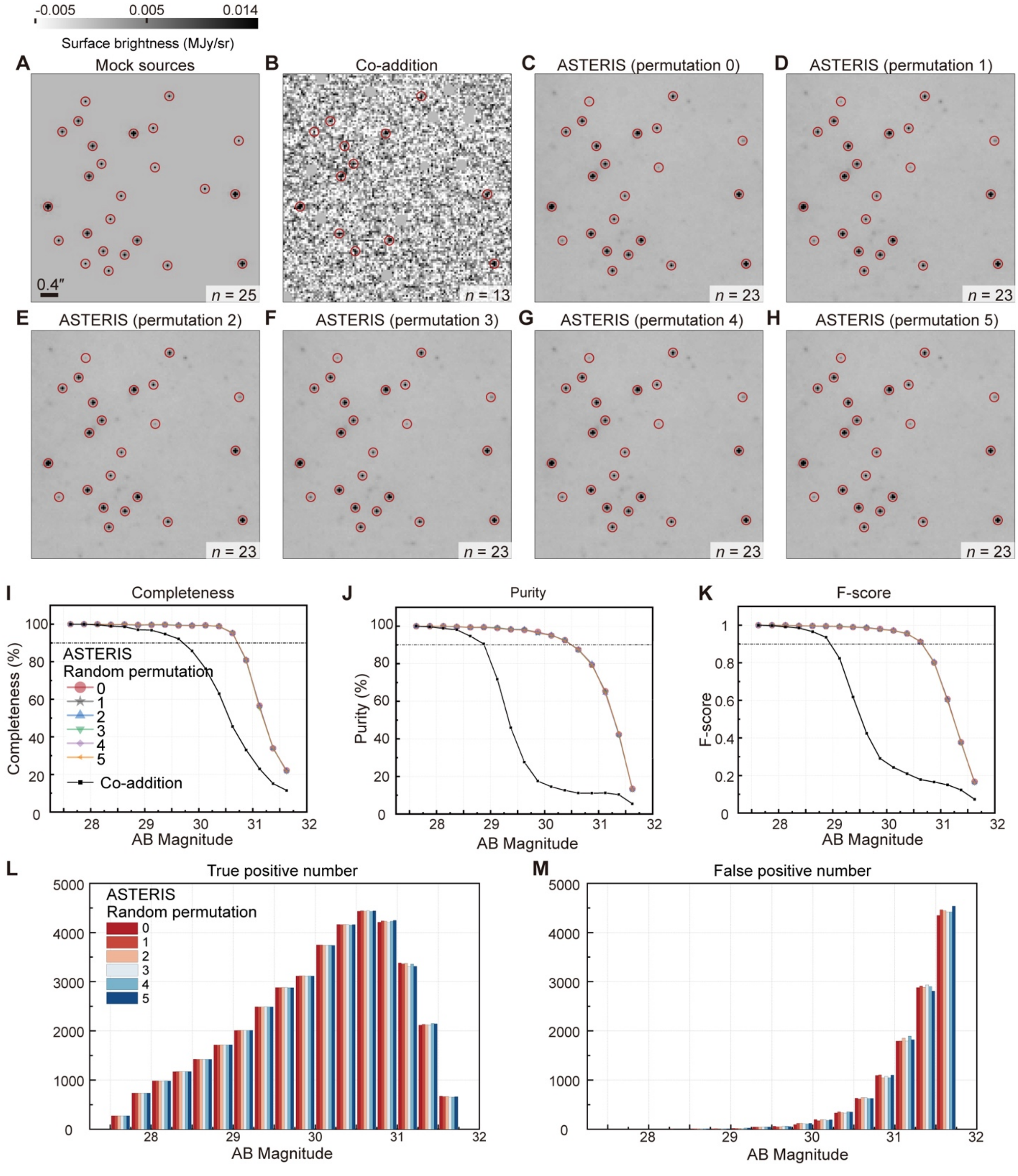

**Fig. S9. Mock tests of ASTERIS robustness to temporal permutation. (A)** Isolated ground truth mock sources same as Fig.2A. (**B**) Co-addition result by averaging eight exposures same as Fig. 2B. (**C**) ASTERIS run on the 8 exposures same as Fig. 2D. (**D** to **H**) Same as panel C, but showing ASTERIS results from five different permutations of the input order for the eight exposures. The number *n* is the same as in Fig. S6. Scale bar, 0.4 arcseconds. **(I)** Comparison of detection completeness from the different input permutations (colored lines, see legend). The 90% completeness level is plotted by a black dashed line. **(J)** Comparisons of detection purity for the different input permutations. The 90% purity level is plotted by a black dashed line. **(K)** Comparison of F-score for the different input permutations. An F-score of 0.9 is shown as a black dashed line. **(E)** Histogram of true positives as a function of apparent magnitude, compared between different input permutations (color, see legend). The number of source detections in each bin is almost identical for the different permutations. **(F)** Same as panel E, but for false positives.

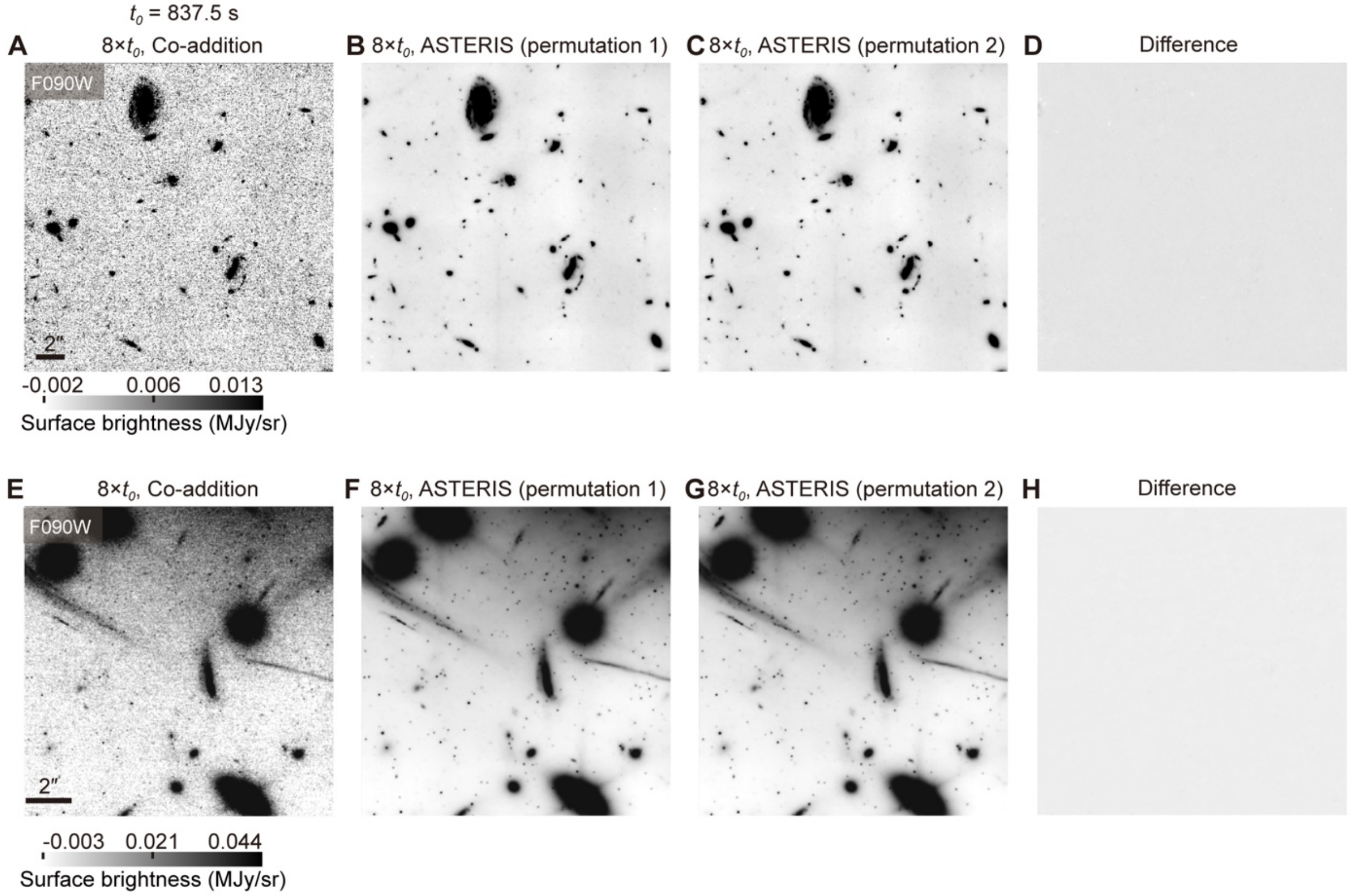

**Fig. S10. Additional tests of ASTERIS robustness to temporal permutation.** Results based on the JWST GLIMPSE program NIRCam F090W imaging. (**A** to **D**) Comparisons of a randomly chosen field. **(A)** Co-addition result of 8×$t_0$ exposures. (**B** to **C**) ASTERIS result of 8×$t_0$ exposures with two different input permutations. **(D)** Difference image between (B) and (C). (**E** to **H**) Same as panels A-D, but for a crowded field in F090W. Scale bars, 2 arcseconds.

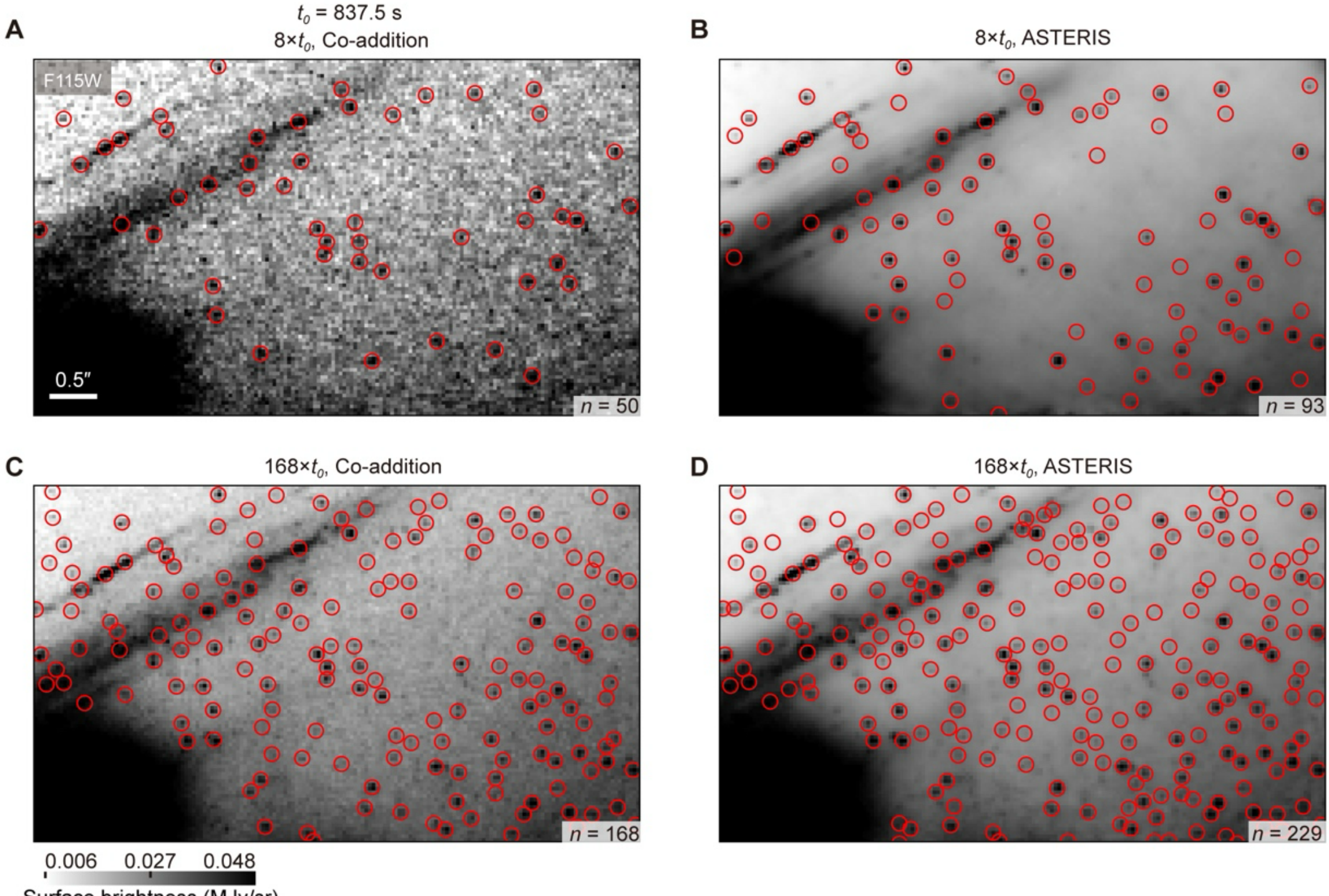

**Fig. S11. Source detection using Source Extractor.** Example of a crowded field from Fig. 3A-D near a bright source from the JWST GLIMPSE program NIRCam F115W imaging. Detected sources are denoted by red circles. (**A**) Co-addition result of 8×$t_0$ exposures. (**B**) ASTERIS result of 8×$t_0$ exposures. (**C**) Co-addition result of 168×$t_0$ exposures, serving as the ground truth for panels A and B. (**D**) ASTERIS result of 168×$t_0$ exposures. Scale bar, 0.5 arcseconds.

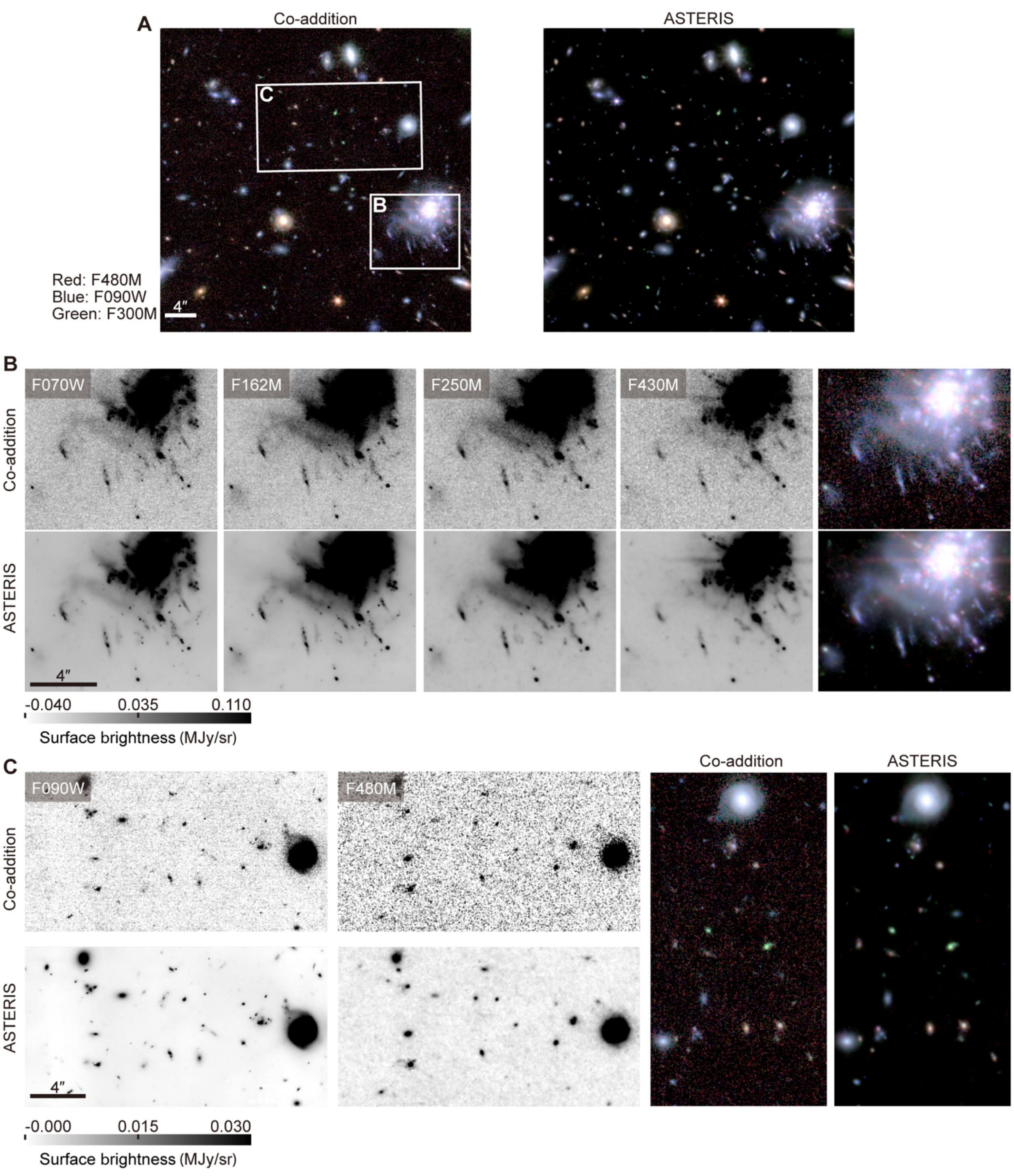

**Fig. S12. Validation of ASTERIS using data from the JWST Mega Science Survey (Program ID 4111). (A)** False-color RGB composites of a region observed by Program 4111, comparing the results from standard co-addition and ASTERIS (blue: F090W, green: F300M, red: F480M for panels A to C). White boxes outline regions shown in panels B and C. **(B)** Zoomed region from panel A. Grayscale images in four bands, and the corresponding color composite. **(C)** Same as panel B, but for only two bands in a different part of panel A. Scale bars, 4 arcseconds.

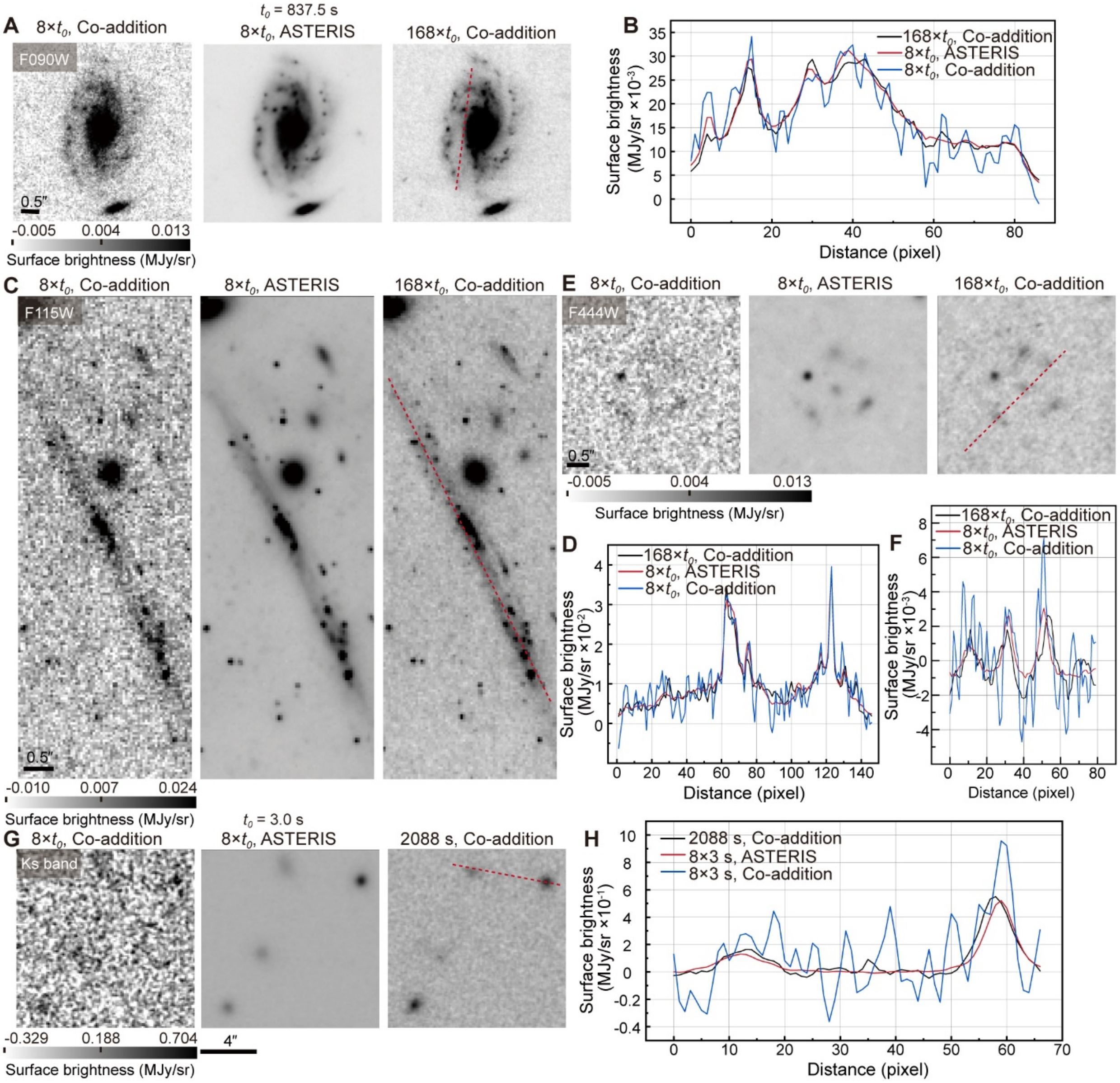

**Fig. S13. Spatial profile comparisons of low-surface-brightness sources.** A comparison between ASTERIS and standard co-addition for the same three regions as in Figure 3. **(A)** The example spiral galaxy from Fig. 3G-I, observed in F090W. The deeper 168-exposure co-addition result serves as the ground-truth reference. **(B)** 1D profile taken along the red dashed line in panel A. Colored lines (see legend) indicate the standard co-addition, ASTERIS and the ground-truth image. **(C-D)** Same as panels A-B, but for the gravitationally lensed arc in Fig. 3J-L, observed in F115W. **(E-F)** Same as panels A-B, but for the group of faint and diffuse galaxies in Fig. 3M-N observed in F444W. Scale bars in panels A, C and E are 0.5 arcseconds. **(G-H)** Similar to panels A-B, but for the Subaru/MOIRCS Ks-band imaging in Fig. 3P-R with $t_0$ = 3 s. Scale bar in panel G is 4 arcseconds.

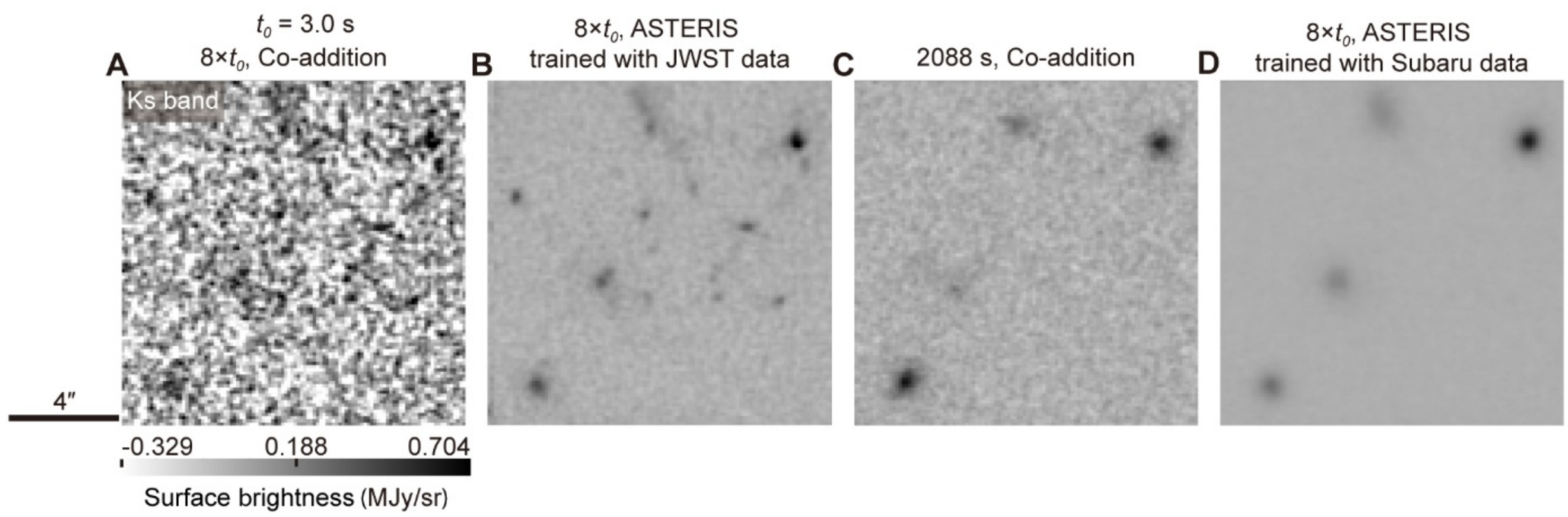

**Fig. S14. Comparison between JWST-trained and Subaru-trained ASTERIS models to Subaru data.** A group of faint sources observed using Subaru MOIRCS in Ks-band, with $t_0$ = 3 s. The images resulting from (**A**) co-addition of 8 × $t_0$ exposures, (**B**) ASTERIS trained with JWST data applied to 8 × $t_0$ exposures, and (**C**) co-addition of 2,088 s exposures, serving as the ground truth for panels A and B. (**D**) Same as Fig. 3Q. ASTERIS trained with JWST data applied to 8 × $t_0$ exposures, which shows less false positive sources than the denoised image in panel B. Scale bars, 4 arcseconds.

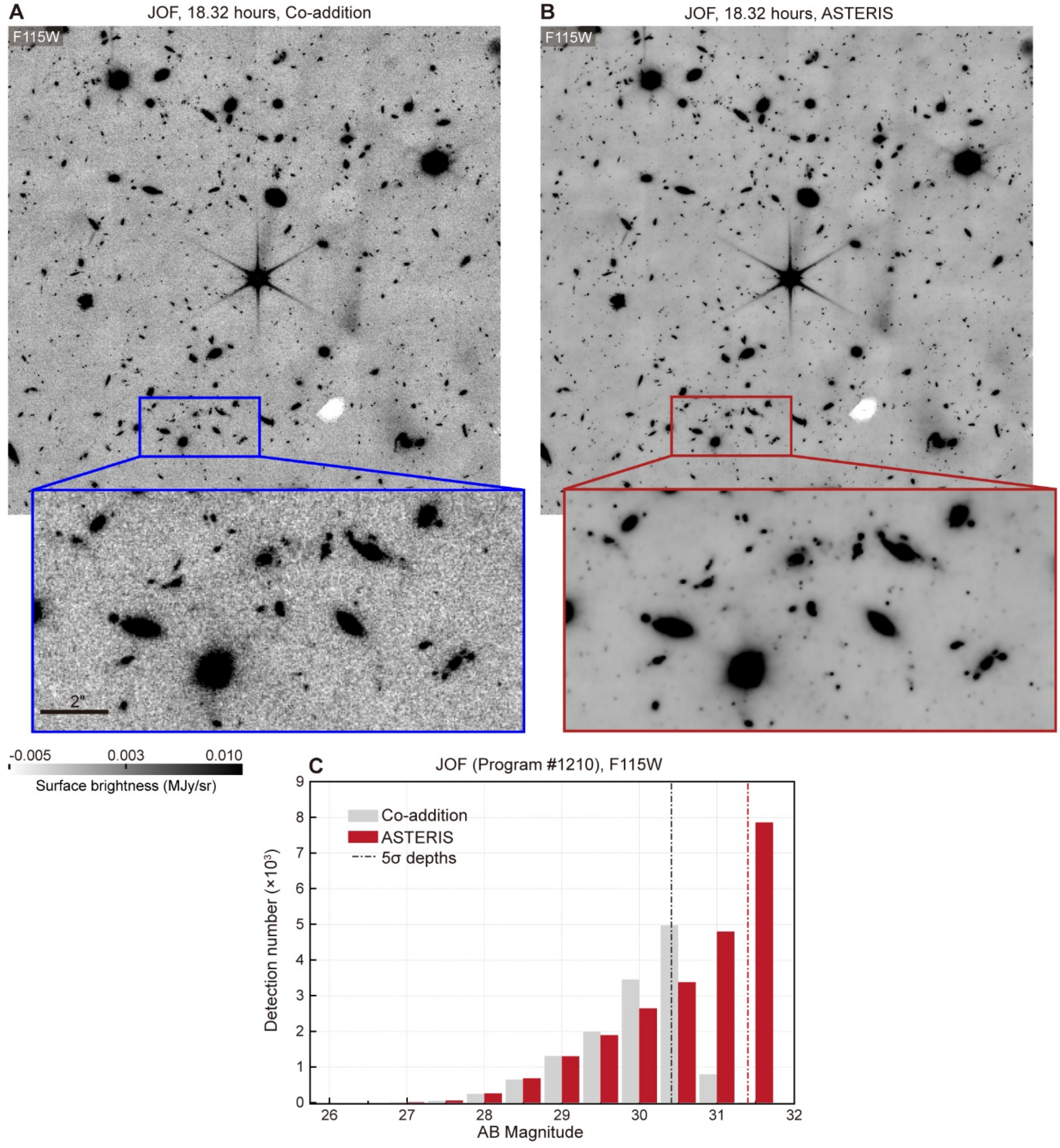

**Fig. S15. Comparison of ASTERIS and co-addition on the JOF field (F115W). (A)** Standard co-addition image, with the blue boxes showing a zoom-in on a representative region. Scale bar, 2 arcseconds. **(B)** ASTERIS-denoised image with a zoom-in on the same region. **(C)** Histogram of the number of detected sources as a function of apparent magnitude for co-addition (gray) and ASTERIS (red). The dashed lines indicate the 5σ depth for each method.

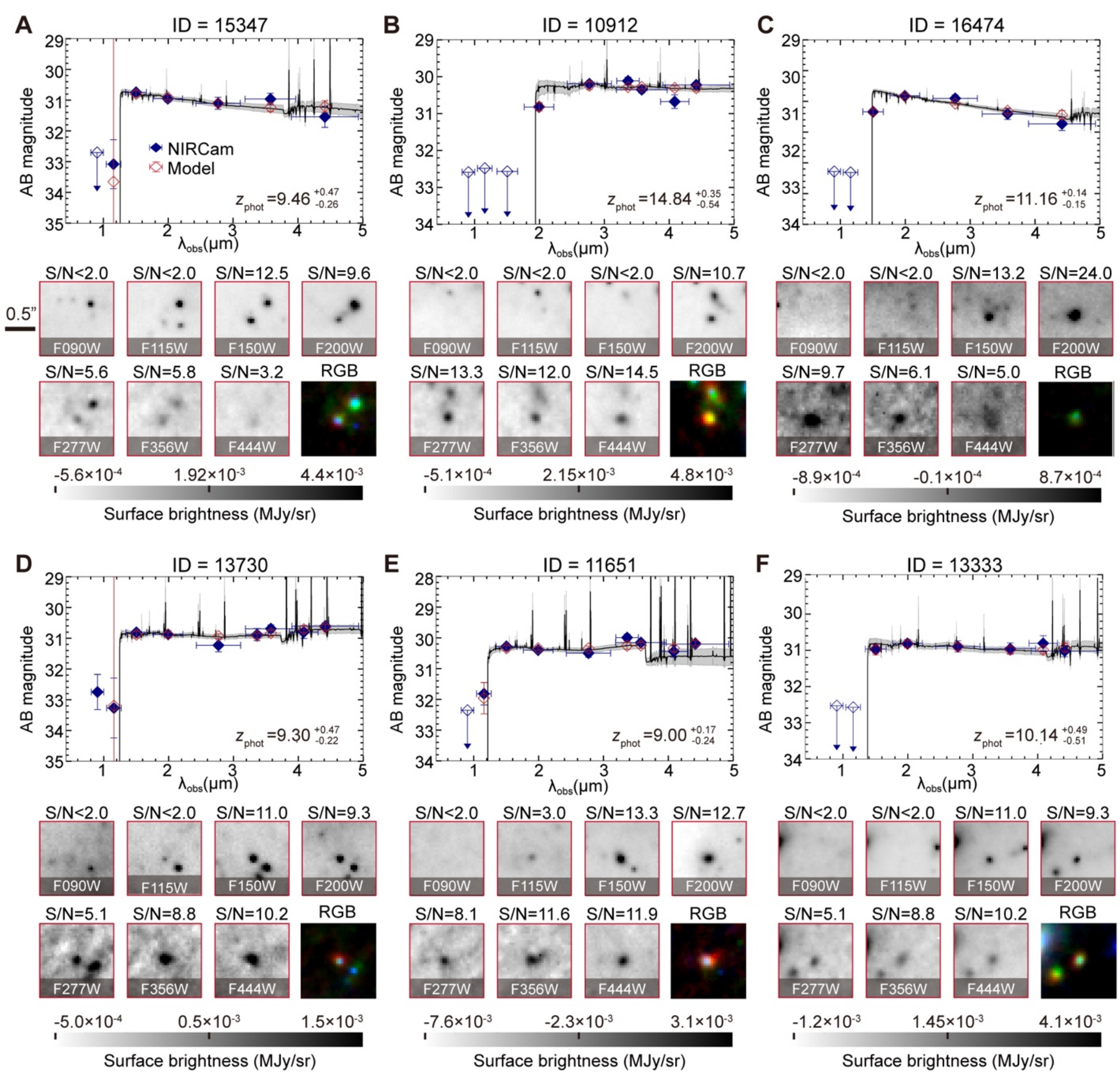

**Fig. S16. Spectral energy distributions and postage stamp images of high-redshift galaxy candidates. (A to E)** Same as Fig. 5B-D, but for five additional high-redshift candidates with their false-color RGB composite images. Blue is F115W + F150W; green is F200W + F277W; red is F356W + F444W. (**F**) Same as Fig. 5B. The scale bar is 0.5 arcseconds.

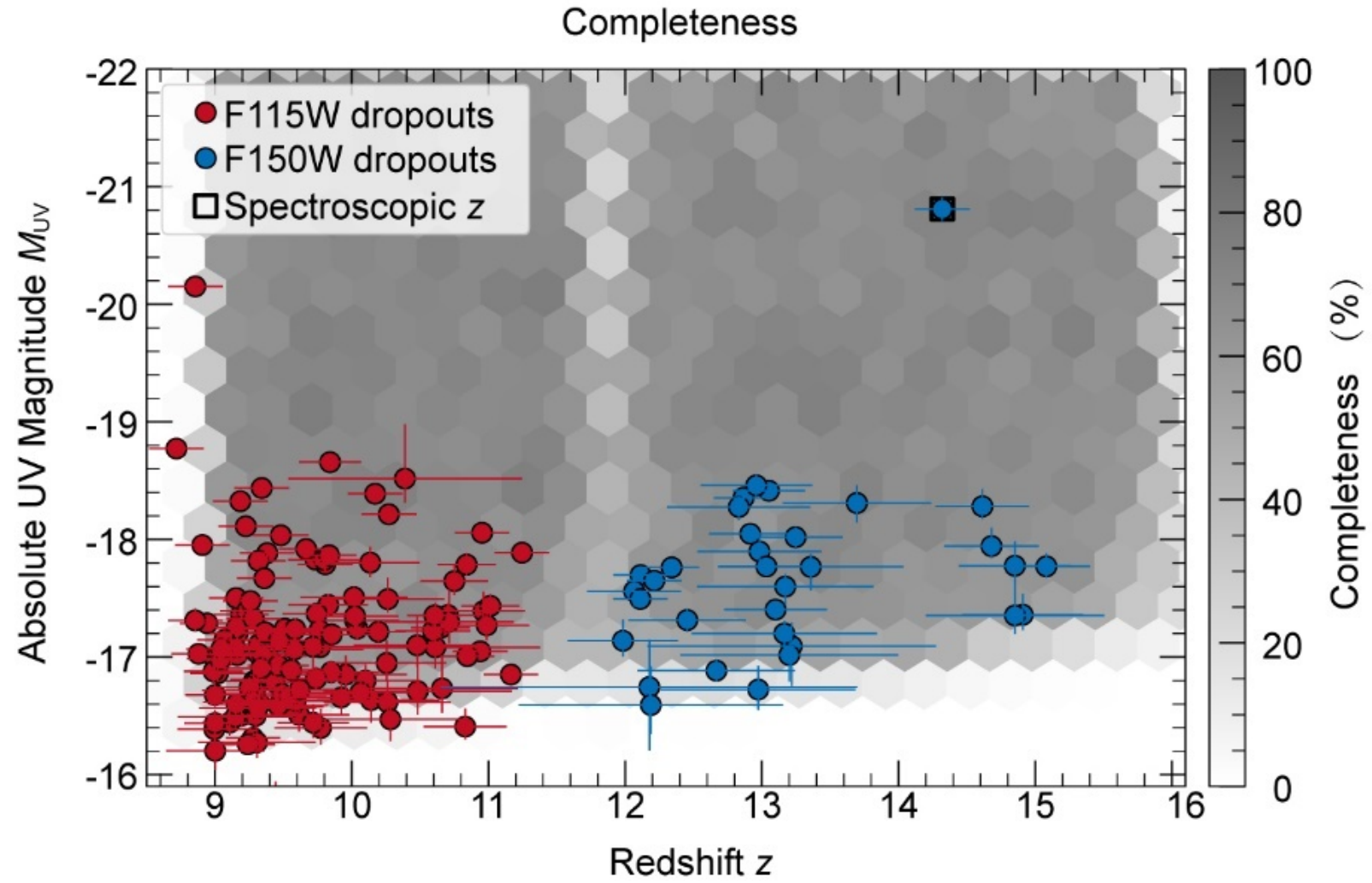

**Fig. S17. The distribution of absolute UV magnitudes and redshifts of high-redshift galaxies.** F115W dropouts (red) and the F150W dropouts (blue) are denoted by dots with error bars indicating 1σ uncertainties. Symbols with black outlines are spectroscopically confirmed (*1*). Grayscale shows the total completeness (color bar) induced by source detection and the high-redshift galaxy selection cuts, evaluated in hexagonal bins.

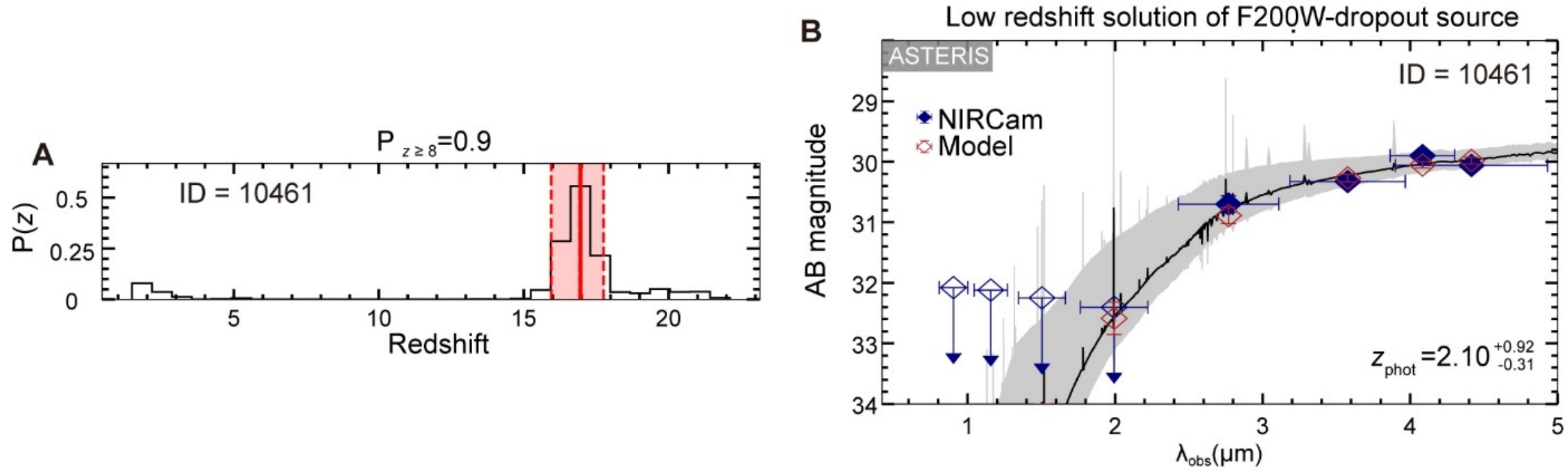

**Fig. S18. Redshift probability distribution and low-redshift solution for a F200W dropout source. (A)** Redshift probability distribution of the F200W dropout source shown in Fig. 5E, which has a high-redshift probability P ($z \geq 8$) = 0.9. **(B)** Same as Fig. 5E but for the alternative low-redshift solution for this F200W dropout source. The low-redshift model is a dusty (dust attenuation in V-band $A_V = 4.4$) dwarf galaxy at $z = 2.1$, with a stellar mass of approximately $2.5\times10^7$ $M_\odot$.

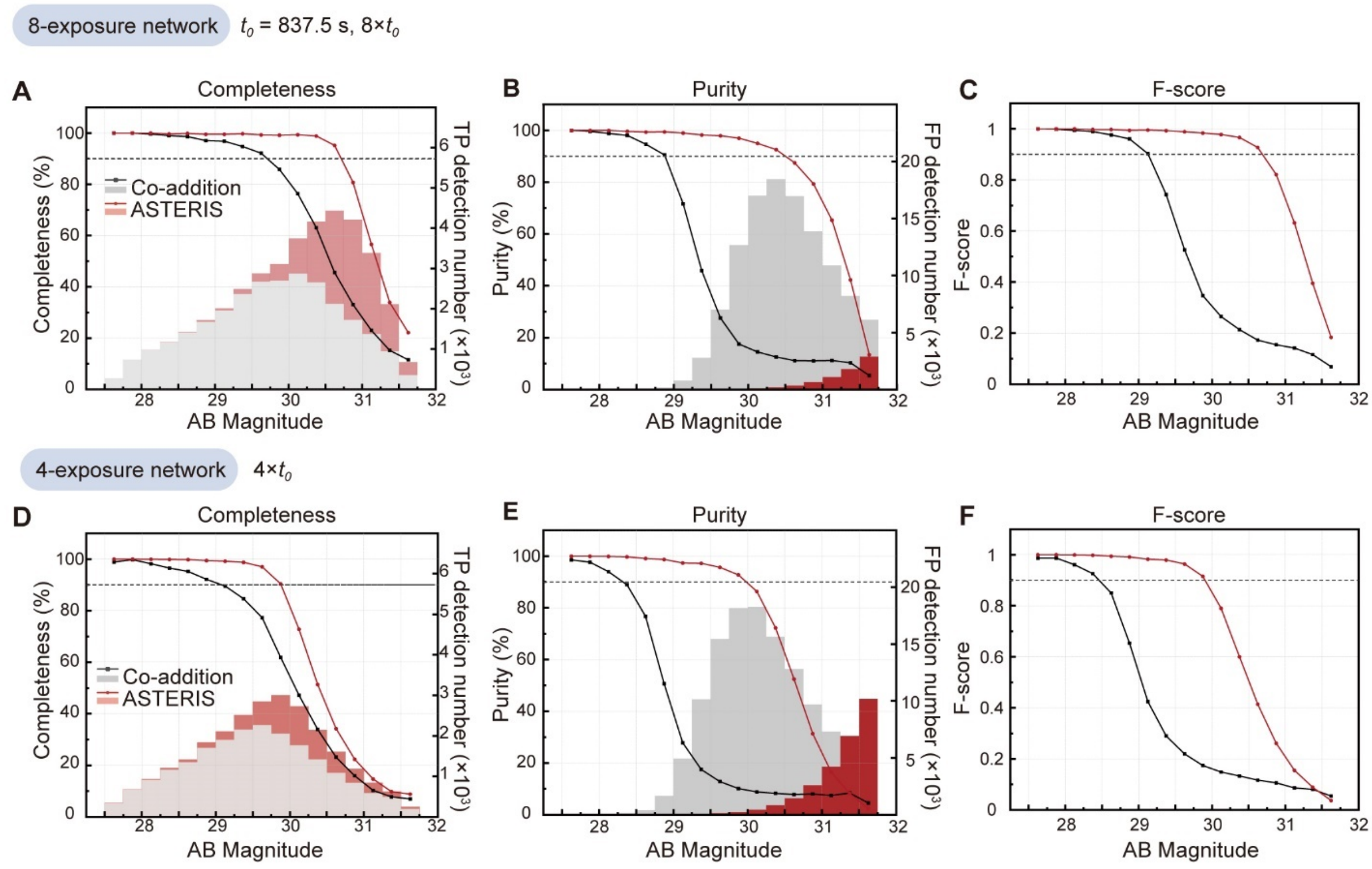

**Fig. S19. Comparison between 8-exposure and 4-exposure ASTERIS models. (A** to **C)** Quantitative evaluation of the 8-exposure ASTERIS model using mock tests. **(A)** Detection completeness and number of true positives, comparing standard co-addition and 8-exposure ASTERIS. The 90% completeness threshold is indicated by a black dashed line. **(B)** Detection purity and number of false positives. The 90% purity threshold is indicated by a black dashed line. **(C)** F-score comparison between the two methods. The 0.9 F-score is shown as a black dashed line. **(D** to **F)** the same as panels A-C, but for the 4-exposure ASTERIS model.

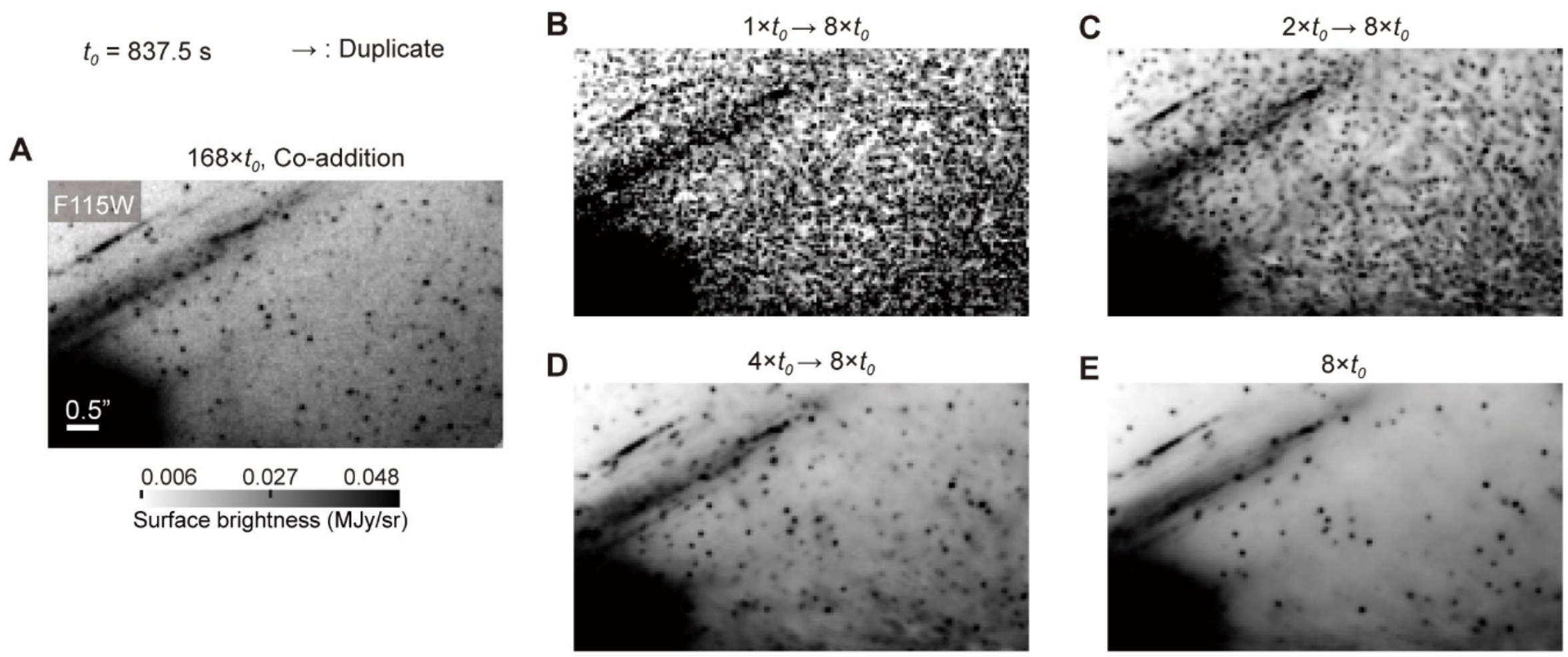

**Fig. S20. Evaluations of artificial input exposure duplication.** Example of a crowded field near a bright source from the JWST GLIMPSE program NIRCam F115W imaging. **(A)** Deep co-addition result using 168×$t_0$ exposures, serving as the ground truth. **(B)** ASTERIS result using 8 duplicated copies of a single 1×$t_0$ exposure, resulting in numerous false positives due to amplification of local noise fluctuations. **(C)** ASTERIS result using four duplicated copies of 2×$t_0$ exposures (8 total exposures). **(D)** ASTERIS result using two duplicated copies of 4×$t_0$ exposures, which still produces a high number of false positives. **(E)** ASTERIS result using 8 independent exposures (no duplication), achieving high detection purity. Scale bars, 0.5 arcseconds.

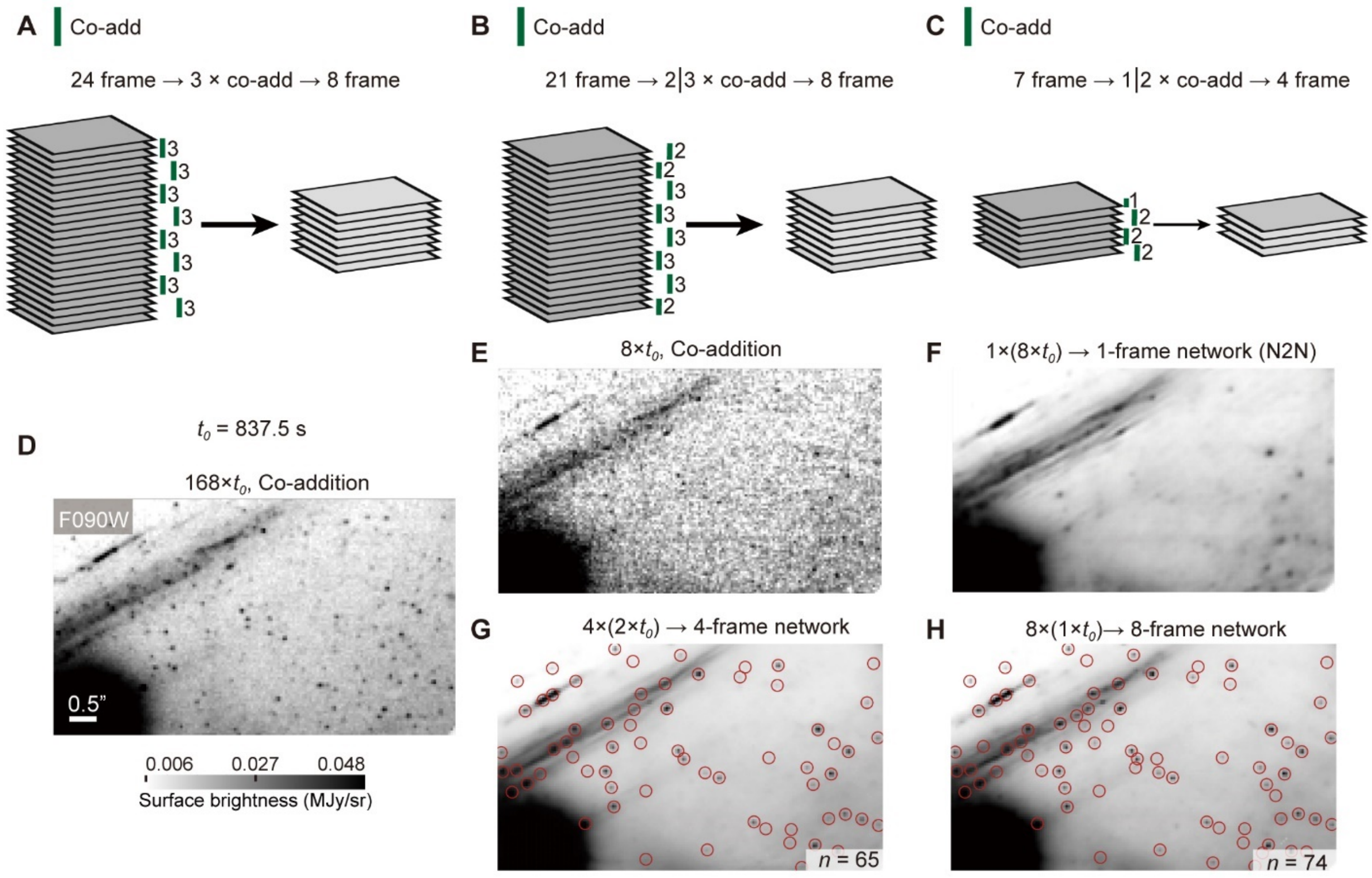

**Fig. S21. Evaluation of three example input configurations for ASTERIS.** (**A**-**C**) When more than $M$ exposures are available, they can be grouped and combined using a subset-measurement co-addition strategy to synthesize the required $M$ input exposures for ASTERIS. To ensure balanced contribution across inputs, each subset is co-added to yield similar depth, minimizing variation in exposure quality. Green ticks and corresponding labels indicate the number of exposures used in each subset. (**D**-**H**) Comparison of the images produced by the subset co-addition strategies in panels A-C, applied to a crowded field near a bright source from the JWST GLIMPSE program NIRCam F090W imaging. (**D**) Deep co-addition result using 168×$t_0$ exposures, serving as the ground truth. (**E**) Standard co-addition result using 8×$t_0$ exposures. (**F**) Noise2Noise result using panel E as input (co-addition before denoising). (**G**) Four-exposure ASTERIS result using 4×(2×$t_0$) inputs, where each input is formed by co-adding two exposures. Completeness is reduced due to information loss during subset co-addition. (**H**) ASTERIS result using 8 independent exposures (no duplication), achieving higher completeness. In panels C-D, $n$ is the number of true-positive sources (red circles) identified using identical Source Extractor parameters. Scale bars, 0.5 arcseconds.

**Table S1. JWST/NIRCam datasets used for training and testing ASTERIS.** Columns include the Program ID, Filters, total Exposure time, and the Number of exposures for each observation program. Detectors in training and testing specify the subsets of NIRCam detectors used for training and testing, respectively. Testing exposures present the total number of exposures allocated for testing. A1-A4, B1-B4 are the short wavelength detectors ($< 2.5$ μm), and A5, B5 are the long wavelength detectors ($\geq 2.5$ μm) (*7*). For Program IDs 1210 and 3215, data from the F335M filter are at the same pointing and therefore stacked for training.

<table>
<thead>
<tr><th>Program ID</th><th>Filters</th><th>Exposure time (hours)</th><th>Number of exposures</th><th>Detectors in training</th><th>Detectors in testing</th><th>Testing exposures</th></tr>
</thead>
<tbody>
<tr><td rowspan="9"><b>3293</b></td><td>F090W</td><td>39.1</td><td>168</td><td>A2-A4,<br>B1-B4</td><td>A1</td><td>8</td></tr>
<tr><td>F115W</td><td>39.1</td><td>168</td><td>A1, A2, A4,<br>B1, B4</td><td>A3,<br>B2, B3</td><td>168</td></tr>
<tr><td>F150W</td><td>22.3</td><td>96</td><td rowspan="2">A1-A4,<br>B1-B4</td><td rowspan="2">–</td><td rowspan="2">–</td></tr>
<tr><td>F200W</td><td>19.5</td><td>96</td></tr>
<tr><td>F277W</td><td>22.3</td><td>96</td><td rowspan="3">A5, B5</td><td rowspan="3">–</td><td rowspan="3">–</td></tr>
<tr><td>F356W</td><td>19.5</td><td>96</td></tr>
<tr><td>F410M</td><td>11.2</td><td>48</td></tr>
<tr><td>F444W</td><td>39.1</td><td>168</td><td>B5</td><td>A5</td><td>8</td></tr>
<tr><td>F480M</td><td>22.3</td><td>96</td><td>A5, B5</td><td>–</td><td>–</td></tr>
<tr><td rowspan="7"><b>1210</b></td><td>F090W</td><td>13.7</td><td>18</td><td rowspan="4">A1-A4,<br>B1-B4</td><td rowspan="7">–</td><td rowspan="7">–</td></tr>
<tr><td>F115W</td><td>18.3</td><td>24</td></tr>
<tr><td>F150W</td><td>13.7</td><td>18</td></tr>
<tr><td>F277W</td><td>11.45</td><td>15</td></tr>
<tr><td>F335M</td><td>6.9</td><td>9</td><td rowspan="3">A5, B5</td></tr>
<tr><td>F410M</td><td>13.7</td><td>18</td></tr>
<tr><td>F444W</td><td>13.7</td><td>18</td></tr>
<tr><td rowspan="6"><b>3215</b></td><td>F162M</td><td>22.9</td><td>30</td><td rowspan="3">A1-A4,<br>B1-B4</td><td rowspan="6">–</td><td rowspan="6">–</td></tr>
<tr><td>F182M</td><td>45.8</td><td>60</td></tr>
<tr><td>F210M</td><td>34.4</td><td>45</td></tr>
<tr><td>F250M</td><td>45.8</td><td>60</td><td rowspan="3">A5, B5</td></tr>
<tr><td>F300M</td><td>34.4</td><td>45</td></tr>
<tr><td>F335M</td><td>22.9</td><td>30</td></tr>
<tr><td rowspan="2"><b>1963</b></td><td>F182M</td><td>7.7</td><td>24</td><td rowspan="2">A1-A4,<br>B1-B4</td><td rowspan="2">–</td><td rowspan="2">–</td></tr>
<tr><td>F210M</td><td>7.7</td><td>24</td></tr>
<tr><td rowspan="13"><b>4111</b></td><td>F070W</td><td>1.14</td><td>8</td><td rowspan="13">–</td><td rowspan="13">A1-A4,<br>B1-B4</td><td>8</td></tr>
<tr><td>F090W</td><td>2.29</td><td>16</td><td>16</td></tr>
<tr><td>F140M</td><td>1.14</td><td>8</td><td>8</td></tr>
<tr><td>F162M</td><td>1.14</td><td>8</td><td>8</td></tr>
<tr><td>F182M</td><td>1.14</td><td>8</td><td>8</td></tr>
<tr><td>F210M</td><td>1.14</td><td>8</td><td>8</td></tr>
<tr><td>F250M</td><td>1.14</td><td>8</td><td>8</td></tr>
<tr><td>F300M</td><td>1.14</td><td>8</td><td>8</td></tr>
<tr><td>F335M</td><td>1.14</td><td>8</td><td>8</td></tr>
<tr><td>F360M</td><td>1.14</td><td>8</td><td>8</td></tr>
<tr><td>F430M</td><td>1.14</td><td>8</td><td>8</td></tr>
<tr><td>F460M</td><td>1.14</td><td>8</td><td>8</td></tr>
<tr><td>F480M</td><td>1.14</td><td>8</td><td>8</td></tr>
</tbody>
</table>

**Table S2. Mock test parameters.**

| Parameter | Value (range) |
|---|---|
| **Program ID** | 3293 |
| **Filter** | F115W |
| **Detector** | A3 |
| **Number of exposures** | 168 |
| **Source-free region size [arcsec$^2$]** | 5.12 × 5.12 |
| **Pixel scale [arcsec]** | 0.04 |
| **Number of mock groups** | 2000 |
| **Number of sources per group** | 25 |
| **Synthetic sources power-law exponent** | 3 |
| **AB mag range [AB mag]** | [27.5, 31.5] |
| **Empirical PSF** | `STPSF` (*59*) |

**Table S3. High-redshift galaxy candidates identified using ASTERIS.** ID is the source number in our catalog; RA is the right ascension; DEC is the declination, the coordinates are referenced to the J2000 equinox; $z_{\mathrm{phot}}$ is the median of the photometric redshift estimated from `BEAGLE`; $M_{\mathrm{UV}}$ is the median rest-frame UV absolute AB magnitude, and 'Drop-out' indicates which filter this source is a dropout. The low-redshift candidate identified by ASTERIS (Fig. 5C-D) is listed at the bottom of this table. A machine-readable version of this table in CSV format is provided in Data S1.

| ID | RA [deg] | DEC [deg] | $z_{\mathrm{phot}}$ | $M_{\mathrm{UV}}$ [mag] | Drop-out |
|---|---|---|---|---|---|
| 40037 | 53.082112 | -27.84208 | 9.17 | -17.2 | F115W |
| 976 | 53.058322 | -27.884859 | 8.86 | -20.2 | F115W |
| 3847 | 53.054154 | -27.893572 | 8.72 | -18.8 | F115W |
| 4324 | 53.050479 | -27.876496 | 9.34 | -18.4 | F115W |
| 6021 | 53.072481 | -27.855353 | 10.19 | -17.2 | F115W |
| 7130 | 53.030788 | -27.897387 | 8.91 | -18.0 | F115W |
| 7336 | 53.097493 | -27.856774 | 9.23 | -18.1 | F115W |
| 7681 | 53.074783 | -27.844886 | 9.48 | -18.0 | F115W |
| 7855 | 53.092072 | -27.852781 | 9.81 | -17.8 | F115W |
| 7888 | 53.067758 | -27.86455 | 9.19 | -18.3 | F115W |
| 8312 | 53.052493 | -27.890993 | 10.38 | -18.5 | F115W |
| 8325 | 53.097476 | -27.856986 | 9.84 | -18.7 | F115W |
| 8513 | 53.016536 | -27.881945 | 10.17 | -18.4 | F115W |
| 8550 | 53.083273 | -27.872176 | 9.19 | -17.4 | F115W |
| 8928 | 53.048874 | -27.890448 | 9.73 | -17.8 | F115W |
| 9044 | 53.085365 | -27.850469 | 9.80 | -17.8 | F115W |
| 9147 | 53.025232 | -27.875202 | 9.38 | -17.9 | F115W |
| 9259 | 53.058743 | -27.88574 | 10.04 | -17.2 | F115W |
| 9394 | 53.076075 | -27.856023 | 10.27 | -18.2 | F115W |
| 9498 | 53.023569 | -27.879833 | 9.32 | -17.8 | F115W |
| 9656 | 53.028783 | -27.903229 | 9.42 | -17.1 | F115W |
| 9779 | 53.049118 | -27.876353 | 9.36 | -17.7 | F115W |
| 9837 | 53.074763 | -27.845422 | 9.67 | -17.9 | F115W |
| 9927 | 53.033006 | -27.871103 | 10.96 | -17.4 | F115W |
| 9964 | 53.054224 | -27.897382 | 10.26 | -17.5 | F115W |
| 10119 | 53.079006 | -27.863592 | 9.83 | -17.4 | F115W |
| 10360 | 53.014799 | -27.885428 | 8.94 | -17.3 | F115W |
| 10428 | 53.06006 | -27.89143 | 11.25 | -17.9 | F115W |
| 10435 | 53.039625 | -27.905555 | 9.28 | -16.8 | F115W |
| 10464 | 53.098716 | -27.860203 | 10.13 | -17.8 | F115W |
| 10750 | 53.060203 | -27.856259 | 9.83 | -17.9 | F115W |
| 10760 | 53.039343 | -27.872243 | 9.31 | -17.0 | F115W |
| 10818 | 53.04765 | -27.871382 | 9.77 | -16.4 | F115W |
| 10835 | 53.072061 | -27.841734 | 10.65 | -17.2 | F115W |
| 10862 | 53.04113 | -27.866054 | 9.15 | -17.5 | F115W |
| 10926 | 53.039794 | -27.906905 | 9.02 | -16.9 | F115W |
| 10967 | 53.082847 | -27.872647 | 10.75 | -17.6 | F115W |
| 11076 | 53.048577 | -27.888751 | 9.15 | -17.2 | F115W |
| 11125 | 53.055733 | -27.879239 | 10.61 | -17.1 | F115W |
| 11326 | 53.041181 | -27.864073 | 9.51 | -17.2 | F115W |
| 11388 | 53.038105 | -27.871414 | 9.31 | -16.8 | F115W |
| 11402 | 53.03826 | -27.872068 | 9.23 | -17.4 | F115W |
| 11422 | 53.102865 | -27.864197 | 10.49 | -17.1 | F115W |
| 11452 | 53.085027 | -27.874119 | 9.11 | -16.4 | F115W |
| 11469 | 53.104561 | -27.858701 | 10.84 | -17.8 | F115W |
| 11516 | 53.052859 | -27.892494 | 10.95 | -18.1 | F115W |
| 11528 | 53.056808 | -27.888986 | 9.74 | -17.4 | F115W |
| 11572 | 53.031709 | -27.903482 | 9.77 | -17.1 | F115W |
| 11651 | 53.08896 | -27.856506 | 9.00 | -17.0 | F115W |
| 11698 | 53.052051 | -27.879842 | 9.95 | -16.8 | F115W |
| 11751 | 53.070893 | -27.861733 | 10.01 | -17.5 | F115W |
| 11790 | 53.070178 | -27.858698 | 9.28 | -17.3 | F115W |
| 11803 | 53.048739 | -27.872818 | 10.59 | -17.2 | F115W |
| 11819 | 53.033259 | -27.882427 | 9.19 | -16.5 | F115W |
| 11820 | 53.090747 | -27.841356 | 10.03 | -17.3 | F115W |
| 12070 | 53.034313 | -27.906936 | 9.32 | -17.1 | F115W |
| 12071 | 53.093087 | -27.864537 | 9.16 | -17.0 | F115W |
| 12123 | 53.059207 | -27.885921 | 9.07 | -17.1 | F115W |
| 12157 | 53.093087 | -27.858781 | 9.32 | -16.6 | F115W |
| 12175 | 53.031184 | -27.901435 | 11.01 | -17.4 | F115W |
| 12192 | 53.081005 | -27.864859 | 9.12 | -17.1 | F115W |
| 12271 | 53.034823 | -27.892532 | 9.38 | -17.0 | F115W |
| 12284 | 53.051485 | -27.88015 | 9.76 | -17.2 | F115W |
| 12400 | 53.076871 | -27.853625 | 9.70 | -17.1 | F115W |
| 12472 | 53.039067 | -27.865064 | 9.26 | -17.5 | F115W |
| 12492 | 53.03139 | -27.872194 | 10.70 | -17.4 | F115W |
| 12517 | 53.067973 | -27.847511 | 10.10 | -16.8 | F115W |
| 12520 | 53.072934 | -27.844141 | 9.00 | -16.2 | F115W |
| 12565 | 53.068289 | -27.851329 | 9.58 | -17.1 | F115W |
| 12791 | 53.025016 | -27.899148 | 10.26 | -17.0 | F115W |
| 12841 | 53.047352 | -27.876659 | 10.02 | -17.3 | F115W |
| 12848 | 53.085481 | -27.848626 | 9.44 | -16.8 | F115W |
| 12849 | 53.076229 | -27.855053 | 10.49 | -16.7 | F115W |
| 13115 | 53.089261 | -27.862192 | 10.99 | -17.3 | F115W |
| 13163 | 53.055651 | -27.880432 | 9.35 | -17.0 | F115W |
| 13180 | 53.075771 | -27.863516 | 8.85 | -17.3 | F115W |
| 13300 | 53.063242 | -27.859205 | 9.00 | -16.4 | F115W |
| 13314 | 53.06317 | -27.850342 | 9.61 | -16.5 | F115W |
| 13333 | 53.075889 | -27.842700 | 10.14 | -16.6 | F115W |
| 13344 | 53.063769 | -27.852484 | 10.72 | -17.3 | F115W |
| 13347 | 53.019686 | -27.88442 | 9.29 | -16.7 | F115W |
| 13489 | 53.043714 | -27.898116 | 9.56 | -16.6 | F115W |
| 13582 | 53.031764 | -27.895955 | 9.36 | -17.2 | F115W |
| 13645 | 53.046071 | -27.879354 | 9.59 | -16.6 | F115W |
| 13699 | 53.072766 | -27.852096 | 9.85 | -17.2 | F115W |
| 13730 | 53.063157 | -27.852532 | 9.30 | -16.5 | F115W |
| 13885 | 53.03291 | -27.908439 | 9.26 | -16.6 | F115W |
| 13914 | 53.101318 | -27.856959 | 9.85 | -16.9 | F115W |
| 13951 | 53.054871 | -27.88609 | 10.66 | -16.7 | F115W |
| 13978 | 53.088213 | -27.859326 | 9.25 | -16.7 | F115W |
| 13983 | 53.057599 | -27.88056 | 9.47 | -16.9 | F115W |
| 14008 | 53.044514 | -27.886452 | 8.99 | -16.9 | F115W |
| 14106 | 53.042201 | -27.876426 | 9.15 | -17.1 | F115W |
| 14158 | 53.060974 | -27.854266 | 9.00 | -16.4 | F115W |
| 14171 | 53.021715 | -27.883518 | 9.55 | -16.9 | F115W |
| 14459 | 53.055481 | -27.882068 | 9.04 | -17.0 | F115W |
| 14509 | 53.027725 | -27.898277 | 9.00 | -16.7 | F115W |
| 14666 | 53.03395 | -27.87498 | 9.28 | -16.3 | F115W |
| 14883 | 53.10324 | -27.855974 | 9.47 | -17.1 | F115W |
| 14918 | 53.05757 | -27.886501 | 10.26 | -16.6 | F115W |
| 14944 | 53.048647 | -27.890656 | 9.72 | -16.4 | F115W |
| 15101 | 53.092912 | -27.84607 | 10.61 | -17.4 | F115W |
| 15152 | 53.024202 | -27.889315 | 9.92 | -16.7 | F115W |
| 15347 | 53.060493 | -27.856172 | 9.46 | -16.6 | F115W |
| 15355 | 53.041968 | -27.868206 | 9.14 | -16.5 | F115W |
| 15634 | 53.050817 | -27.885667 | 9.74 | -16.8 | F115W |
| 15660 | 53.100503 | -27.849021 | 9.30 | -16.3 | F115W |
| 16023 | 53.059426 | -27.854761 | 9.27 | -16.6 | F115W |

| 16037 | 53.03872 | -27.871565 | 10.29 | -16.5 | F115W |
|---|---|---|---|---|---|
| 16056 | 53.029997 | -27.878509 | 9.05 | -17.1 | F115W |
| 16057 | 53.042712 | -27.869748 | 9.24 | -16.3 | F115W |
| 16196 | 53.051223 | -27.897506 | 9.16 | -16.6 | F115W |
| 16289 | 53.032589 | -27.905758 | 9.44 | -15.7 | F115W |
| 16439 | 53.033681 | -27.894577 | 9.50 | -16.7 | F115W |
| 16452 | 53.031858 | -27.895427 | 10.83 | -16.4 | F115W |
| 16474 | 53.026999 | -27.898086 | 11.16 | -16.9 | F115W |
| 17387 | 53.050669 | -27.883014 | 10.94 | -17.0 | F115W |
| 17557 | 53.042163 | -27.879839 | 9.34 | -16.9 | F115W |
| 17634 | 53.040206 | -27.87669 | 9.54 | -16.7 | F115W |
| 17825 | 53.022986 | -27.883021 | 9.62 | -16.7 | F115W |
| 17934 | 53.073198 | -27.84167 | 10.06 | -16.7 | F115W |
| 20342 | 53.098099 | -27.850742 | 9.31 | -16.6 | F115W |
| 35703 | 53.072878 | -27.861566 | 9.59 | -17.2 | F115W |
| 35706 | 53.058724 | -27.885673 | 10.84 | -17.0 | F115W |
| 35712 | 53.064121 | -27.862256 | 8.88 | -17.0 | F115W |
| 40044 | 53.040165 | -27.876027 | 12.11 | -17.7 | F150W |
| 40054 | 53.082649 | -27.864818 | 13.01 | -17.9 | F150W |
| 2022 | 53.082937 | -27.855633 | 14.32 | -20.8 | F150W |
| 7577 | 53.080527 | -27.870979 | 11.99 | -17.1 | F150W |
| 7707 | 53.046055 | -27.869631 | 13.26 | -18.0 | F150W |
| 8135 | 53.02869 | -27.893013 | 12.87 | -18.3 | F150W |
| 8227 | 53.047445 | -27.872083 | 12.76 | -18.3 | F150W |
| 8477 | 53.045239 | -27.870074 | 12.33 | -17.8 | F150W |
| 8528 | 53.064756 | -27.890234 | 13.05 | -18.4 | F150W |
| 8578 | 53.07934 | -27.871847 | 14.64 | -18.3 | F150W |
| 8916 | 53.084681 | -27.866661 | 12.08 | -17.6 | F150W |
| 9481 | 53.043575 | -27.872944 | 13.79 | -18.3 | F150W |
| 10291 | 53.042921 | -27.870455 | 12.94 | -18.5 | F150W |
| 10643 | 53.026742 | -27.894634 | 13.28 | -17.8 | F150W |
| 10912 | 53.10763 | -27.860128 | 14.84 | -17.8 | F150W |
| 11463 | 53.086693 | -27.872383 | 15.07 | -17.8 | F150W |
| 11598 | 53.075344 | -27.870049 | 14.63 | -17.9 | F150W |
| 11760 | 53.092205 | -27.84587 | 12.93 | -18.0 | F150W |
| 12016 | 53.104693 | -27.861868 | 12.98 | -17.8 | F150W |
| 12334 | 53.029782 | -27.892389 | 13.11 | -17.4 | F150W |
| 12512 | 53.068635 | -27.846787 | 13.10 | -17.6 | F150W |
| 13484 | 53.045084 | -27.897909 | 12.27 | -16.8 | F150W |

| 13495 | 53.036131 | -27.902974 | 14.88 | -17.4 | F150W |
|---|---|---|---|---|---|
| 13720 | 53.075939 | -27.842576 | 12.96 | -17.1 | F150W |
| 14642 | 53.040176 | -27.876021 | 12.12 | -17.5 | F150W |
| 14984 | 53.070488 | -27.871587 | 14.85 | -17.4 | F150W |
| 15124 | 53.082524 | -27.851093 | 12.21 | -17.7 | F150W |
| 15153 | 53.027678 | -27.886677 | 13.25 | -17.2 | F150W |
| 15796 | 53.032744 | -27.886077 | 12.24 | -16.6 | F150W |
| 16563 | 53.074614 | -27.860042 | 12.94 | -16.7 | F150W |
| 17109 | 53.075759 | -27.877938 | 13.2 | -17.0 | F150W |
| 17782 | 53.064374 | -27.850764 | 12.65 | -16.9 | F150W |
| 35709 | 53.042823 | -27.87177 | 12.47 | -17.3 | F150W |
| 10461 | 53.041949 | -27.900857 | 17.08 | -17.3 | F200W |
| 14307 | 53.102727 | -27.86185 | 18.42 | -17.3 | F200W |
| 16368 | 53.097803 | -27.853101 | 18.54 | -17.0 | F200W |
| 20812 | 53.029207 | -27.887505 | 21.04 | -16.9 | F200W |
| 40011 | 53.019108 | -27.879895 | 2.08 | -11.1 | None |

Q. Ha, P. Haderlein, A. Hagedorn, K. Hainline, C. Haley, M. Hami, F. Hamilton, H. Hammel, C. Hansen, T. Harkins, M. Harr, J. Hart, Q. Hart, G. Hartig, R. Hashimoto, S. Haskins, W. Hathaway, K. Havey, B. Hayden, K. Hecht, C. Heller-Boyer, C. Henriques, A. Henry, K. Hermann, S. Hernandez, B. Hesman, B. Hicks, B. Hilbert, D. Hines, M. Hoffman, S. Holfeltz, B. J. Holler, J. Hoppa, K. Hott, J. M. Howard, R. Howard, A. Hunter, D. Hunter, B. Hurst, B. Husemann, L. Hustak, L. Ilinca Ignat, G. Illingworth, S. Irish, W. Jackson, A. Jahromi, P. Jakobsen, L. A. James, B. James, W. Januszewski, A. Jenkins, H. Jirdeh, P. Johnson, T. Johnson, V. Jones, R. Jones, D. Jones, O. Jones, I. Jordan, M. Jordan, S. Jurczyk, A. Jurling, C. Kaleida, P. Kalmanson, J. Kammerer, H. Kang, S.-H. Kao, D. Karakla, P. Kavanagh, D. Kelly, S. Kendrew, H. Kennedy, D. Kenny, R. Keski-kuha, C. Keyes, R. Kidwell, W. Kinzel, J. Kirk, M. Kirkpatrick, D. Kirshenblat, P. Klaassen, B. Knapp, J. Scott Knight, P. Knollenberg, R. Koehler, A. Koekemoer, A. Kovacs, T. Kulp, N. Kumari, M. Kyprianou, S. La Massa, A. Labador, A. Labiano, P.-O. Lagage, C.-P. Lajoie, M. Lallo, M. Lam, T. Lamb, S. Lambros, R. Lampenfield, J. Langston, K. Larson, D. Law, J. Lawrence, D. Lee, J. Leisenring, K. Lepo, M. Leveille, N. Levenson, M. Levine, Z. Levy, D. Lewis, H. Lewis, M. Libralato, P. Lightsey, M. Link, L. Liu, A. Lo, A. Lockwood, R. Logue, C. Long, D. Long, C. Loomis, M. Lopez-Caniego, J. Lorenzo Alvarez, J. Love-Pruitt, A. Lucy, N. Luetzgendorf, P. Maghami, R. Maiolino, M. Major, S. Malla, E. Malumuth, E. Manjavacas, C. Mannfolk, A. Marrione, A. Marston, A. Martel, M. Maschmann, G. Masci, M. Masciarelli, M. Maszkiewicz, J. Mather, K. McKenzie, B. McLean, M. McMaster, K. Melbourne, M. Meléndez, M. Menzel, K. Merz, M. Meyett, L. Meza, C. Miskey, K. Misselt, C. Moller, J. Morrison, E. Morse, H. Moseley, G. Mosier, M. Mountain, J. Mueckay, M. Mueller, S. Mullally, J. Murphy, K. Murray, C. Murray, D. Mustelier, J. Muzerolle, M. Mycroft, R. Myers, K. Myrick, S. Nanavati, E. Nance, O. Nayak, B. Naylor, E. Nelan, B. Nickson, A. Nielson, M. Nieto-Santisteban, N. Nikolov, A. Noriega-Crespo, B. O'Shaughnessy, B. O'Sullivan, W. Ochs, P. Ogle, B. Oleszczuk, J. Olmsted, S. Osborne, R. Ottens, B. Owens, C. Pacifici, A. Pagan, J. Page, S. Park, K. Parrish, P. Patapis, L. Paul, T. Pauly, C. Pavlovsky, A. Pedder, M. Peek, M. Pena-Guerrero, K. Penanen, Y. Perez, M. Perna, B. Perriello, K. Phillips, M. Pietraszkiewicz, J.-P. Pinaud, N. Pirzkal, J. Pitman, A. Piwowar, V. Platais, D. Player, R. Plesha, J. Pollizi, E. Polster, K. Pontoppidan, B. Porterfield, C. Proffitt, L. Pueyo, C. Pulliam, B. Quirt, I. Quispe Neira, R. Ramos Alarcon, L. Ramsay, G. Rapp, R. Rapp, B. Rauscher, S. Ravindranath, T. Rawle, M. Regan, T. A. Reichard, C. Reis, M. E. Ressler, A. Rest, P. Reynolds, T. Rhue, K. Richon, E. Rickman, M. Ridgaway, C. Ritchie, H.-W. Rix, M. Robberto, G. Robinson, M. Robinson, O. Robinson, F. Rock, D. Rodriguez, B. Rodriguez Del Pino, T. Roellig, S. Rohrbach, A. Roman, F. Romelfanger, P. Rose, A. Roteliuk, M. Roth, B. Rothwell, N. Rowlands, A. Roy, P. Royer, P. Royle, C. Rui, P. Rumler, J. Runnels, M. Russ, Z. Rustamkulov, G. Ryden, H. Ryer, M. Sabata, D. Sabatke, E. Sabbi, B. Samuelson, B. Sapp, B. Sappington, B. Sargent, A. Sauer, S. Scheithauer, E. Schlawin, J. Schlitz, T. Schmitz, A. Schneider, J. Schreiber, V. Schulze, R. Schwab, J. Scott, K. Sembach, C. Shanahan, B. Shaughnessy, R. Shaw, N. Shawger, C. Shay, E. Sheehan, S. Shen, A. Sherman, B. Shiao, H.-Y. Shih, I. Shivaei, M. Sienkiewicz, D. Sing, M. Sirianni, A. Sivaramakrishnan, J. Skipper, G. C. Sloan, C. Slocum, S. Slowinski, E. Smith, E. Smith, D. Smith, C. Smith, G. Snyder, W. Soh, S. Tony Sohn, C. Soto, R. Spencer, S. Stallcup, J. Stansberry, C. Starr, E. Starr, A. Stewart, M. Stiavelli, A. Straughn, D. Strickland, J. Stys, F. Summers, F. Sun, B. Sunnquist, D. Swade, M. Swam, R. Swaters, R. Swoish, J. M. Taylor, R. Taylor, M. Te Plate, M. Tea, K. Teague, R.